\documentclass[journal]{vgtc}                
\ifpdf
  \pdfoutput=1\relax                   
  \usepackage{graphicx}                
  \DeclareGraphicsExtensions{.pdf,.png,.jpg,.jpeg} 
\else
  \usepackage{graphicx}                
  \DeclareGraphicsExtensions{.eps}     
\fi%

\graphicspath{{figures/}{pictures/}{images/}{./}} 

\usepackage{microtype}                 
\PassOptionsToPackage{warn}{textcomp}  
\usepackage{textcomp}                  
\usepackage{mathptmx}                  
\usepackage{times}                     
\usepackage{cite}                      
\usepackage{tabu}                      
\usepackage{booktabs}                  

\usepackage{algorithm}
\usepackage{algorithmic}
\usepackage{float} 
\usepackage {color}
\usepackage{overpic}
\usepackage{booktabs}
\usepackage{multirow}
\usepackage{array}
\usepackage{url}

\onlineid{0}

\vgtccategory{Research}
\vgtcpapertype{please specify}

\title{Incorporation of Human Knowledge into Data Embeddings to Improve Pattern Significance and Interpretability}

\author{Jie Li and Chun-qi Zhou}
 \authorfooter{
 \item
  Jie Li and Chun-qi Zhou are with College of Intelligence and Computing, Tianjin University. Jie Li is the corresponding author. E-mail: jie.li@tju.edu.cn.
}

\abstract{Embedding is a common technique for analyzing multi-dimensional data. However, the embedding projection cannot always form significant and interpretable visual structures that foreshadow underlying data patterns. We propose an approach that incorporates human knowledge into data embeddings to improve pattern significance and interpretability. The core idea is (1) externalizing tacit human knowledge as explicit sample labels and (2) adding a classification loss in the embedding network to encode samples' classes. The approach pulls samples of the same class with similar data features closer in the projection, leading to more compact (significant) and class-consistent (interpretable) visual structures. We give an embedding network with a customized classification loss to implement the idea and integrate the network into a visualization system to form a workflow that supports flexible class creation and pattern exploration. Patterns found on open datasets in case studies, subjects' performance in a user study, and quantitative experiment results illustrate the general usability and effectiveness of the approach.%
} 


\CCScatlist{ 
 \CCScat{K.6.1}{Management of Computing and Information Systems}%
{Project and People Management}{Life Cycle};
 \CCScat{K.7.m}{The Computing Profession}{Miscellaneous}{Ethics}
}

\teaser{
  \centering
    
  \includegraphics[width=\linewidth]{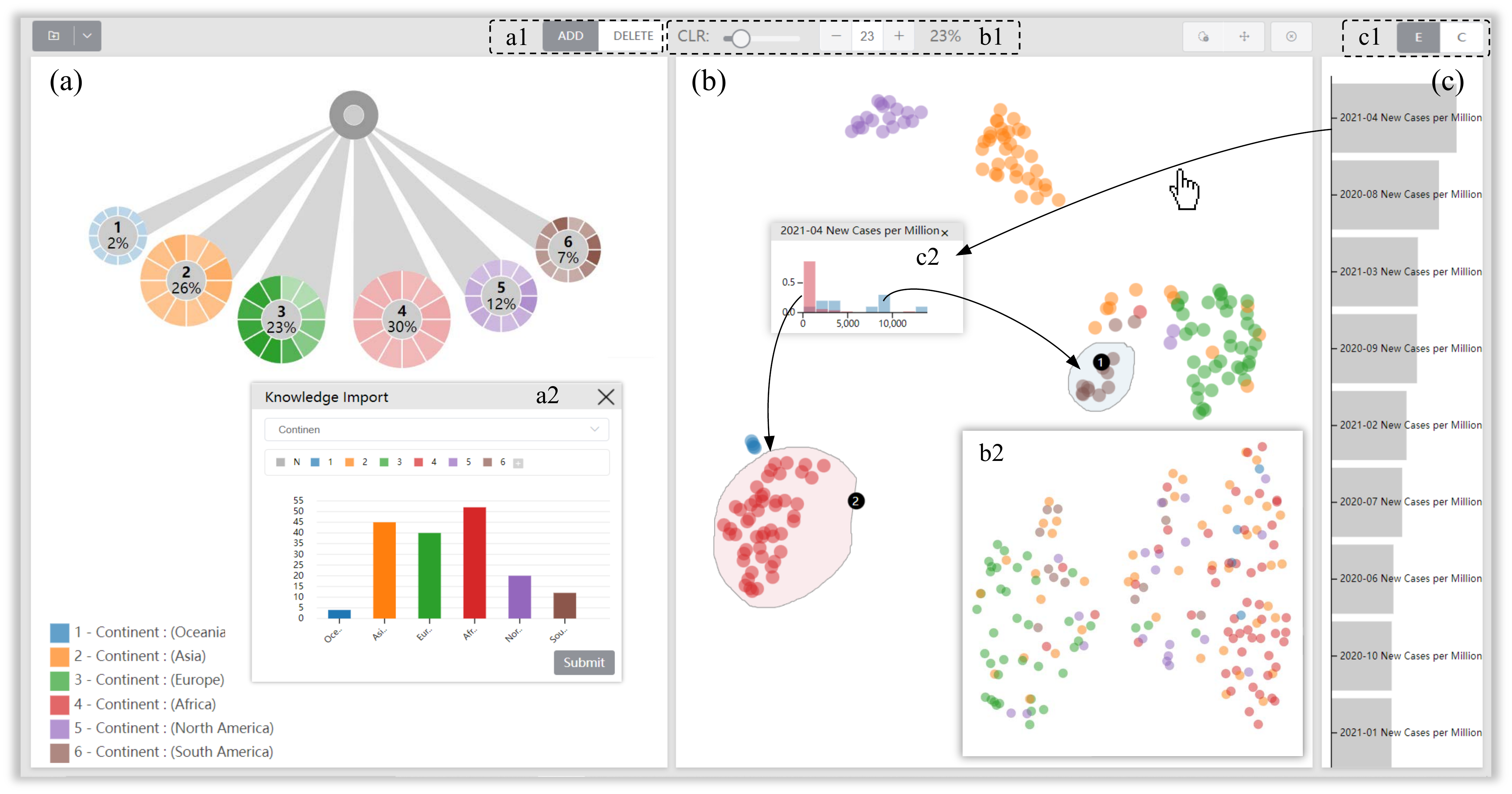}
  \vspace{-0.25in}
  \caption{The visualization system interface that consists of three components: (a) Knowledge Editor (KE), (b) Data Projector (DP), and (c) Pattern Explainer (PE). The analyst is analyzing the Covid-19 dataset\cite{covid_19} that consists of 173 sequences, each recording the number of monthly-confirmed cases (per million people) from May 2020 to April 2021 of a country. The analyst creates six country groups (see colored tree nodes) according to their knowledge using (a) and incorporates the class information into the embeddings of countries to generate the projection with clear and class-consistent visual structures that foreshadow underlying data patterns. (b2) The embedding projection without knowledge where does not exist visual structure.}
  \label{fig6}
}

\begin{document}


\firstsection{Introduction}
\maketitle
\begin{figure}[htbp]
\centering
 \includegraphics[width=\linewidth]{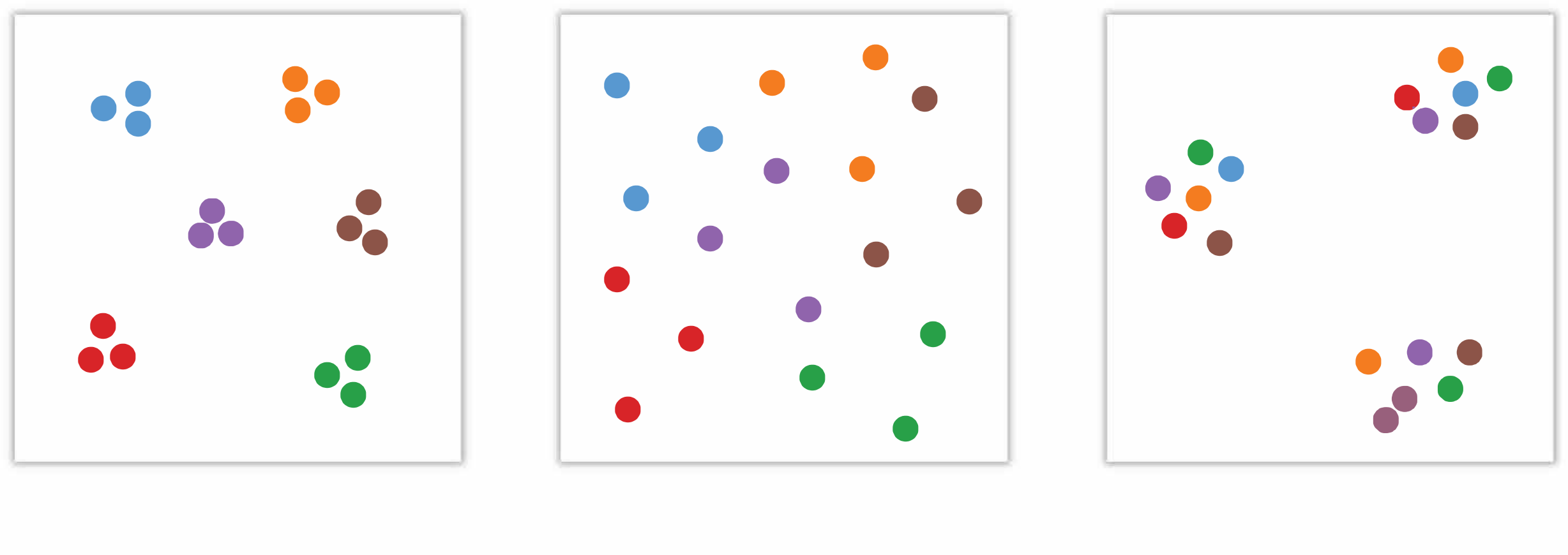}
  \put(-220,5){\small $(a)$}
  \put(-130,5){\small $(b)$}
  \put(-45,5){\small $(c)$}
 \vspace{-0.18in}
\caption{Exploring data patterns in the projection. The color of each sample indicates its value on an attribute. (a) An ideal projection with clear and pure visual structures. 
(b-c) Two negative situations. Visual structures are not compact enough or consist of samples with varying attribute values.}
\label{fig1}
\vspace{-0.1in}
\end{figure}

Embedding is a common technique for analyzing multi-dimensional data. Most embedding networks follow a self-supervised framework to construct a latent space in which the distance between two samples reflects their similarity in data features. Analysts thus can explore data patterns in the embedding projection (abbreviated as projection below). They typically search for special visual structures, e.g., outliers and clusters, and (2) explain patterns of each visual structure by observing whether included samples have any commonality, e.g., similar values on an attribute, as in Figure \ref{fig1}(a).

However, visual structures in embedding projection are not always clear and easy to explain. First, samples with similar data features (e.g., attribute values), although adjacent to each other, may not form compact visual structures, as in Figure \ref{fig1}(b). Therefore, analysts cannot determine reasonable selection boundaries, leading to missing or misidentifying data patterns (\textbf{significance}). Second, samples within a visual structure vary widely in data features, as in Figure \ref{fig1}(c). Analysts thus may consider the visual structure randomly formed and meaningless (\textbf{interpretability}). Most real-world datasets have more or less the two problems, making discovering and understanding data patterns from the projection challenges.

This paper presents an approach that leverages human knowledge to improve pattern significance and interpretability in the embedding projection. The approach involves two key steps. First, we externalize tacit human knowledge as explicit sample labels, arguing that samples with the same label should involve similar data patterns (\textbf{Knowledge Externalization}). Second, we make the embedding network simultaneously encode classes and data features of samples (\textbf{Knowledge Incorporation}). The two steps can pull samples of the same class with similar data features closer in the projection, leading to more compact (\textbf{significant}) and class-consistent (\textbf{interpretable}) visual structures, thus resolving the two problems.

We propose an embedding network to achieve the two steps. The core idea is to equip the network with a reconstruction loss and a classification loss to make the final embeddings reflect sample similarity in both data features and classes. Our embedding network differs from existing supervised embedding networks \cite{le2018supervised} in two aspects. Specifically, our network (1) utilizes a customized classification loss, bringing quick convergence speed and promoting visual structure formation in the projection, and (2) does not have an independent encoder and makes embeddings as the input. Thus, it has high embedding efficiency, conducting model training and sample embedding simultaneously.

We integrate the embedding network into a visualization system to achieve a knowledge-based exploratory workflow. Specifically, the system allows analysts to externalize and refine knowledge on any attribute (\textbf{{Knowledge Editing}}), flexibly select visual structures in the projection established using the embedding network (\textbf{Embedding Exploration}), and analyze the factors leading to the formation of the selected visual structures (\textbf{Pattern Explanation}). The integrated visual and quantitative techniques provide intuitive references, simplify analysts' operations, and improve their confidence in the results.

We evaluate our approach through a series of experiments. Patterns on a synthetic dataset and two open real-world datasets demonstrate the system's usability. We further compare subjects' performance using the system to complete the pre-determined analysis tasks with and without knowledge incorporation. We find that incorporating knowledge into embeddings reduces subjects' completing time and operations since clear and class-consistent visual structures avoid blind sample selections. We finally demonstrate the two advantages of the embedding network through a quantitative experiment.

In short, our main contribution is a novel visual analytics approach that fuses human knowledge and machine intelligence via embedding for efficient pattern discovery. The approach involves three sub-contributions, i.e., (1) an embedding network that achieves the knowledge incorporation, (2) a visualization system that integrates (1) to form a knowledge-based exploratory workflow, and (3) experiments that prove the effectiveness and usability of the approach.

We organize the remaining parts as follows. Section \ref{Related Work} reviews relevant studies. After describing the general idea in Section \ref{Basic Idea}, we introduce the embedding network and the visualization system in Sections \ref{Embedding Network} and \ref{Visualization System}. Sections \ref{Case Study}-\ref{Quantitative experiments} discuss the experiment results. We finally conclude the paper in Section \ref{Conclusion and Future works}.

\section{Related Work}
\label{Related Work}
We review relevant works from the following two aspects.

\subsection{Embedding-based Data Exploration}
As a statistics-based embedding technique, dimensionality reduction (DR) is typical in multi-dimensional data analysis. There are many classic DR algorithms with different principles, such as PCA \cite{wold1987principal}, MDS \cite{borg2005modern}, LLE \cite{roweis2000nonlinear}, t-SNE \cite{van2008visualizing}, UMAP \cite{mcinnes2018umap}, etc. Many recent DR algorithms implement special projection effects via customized objective functions \cite{espadoto2019toward, fujiwara2019incremental, wang2017perception}. These algorithms do not support the incorporation of class information. LDA \cite{balakrishnama1998linear} is a supervised DR algorithm. However, it only captures linear relationships between dimensions and requires the embedding length not to exceed the number of classes, affecting its applicability in more scenarios. 

With the success of deep learning in real scenarios, neural network-based embedding techniques gradually became popular. AutoEncoder \cite{vincent2008extracting, ng2011sparse} is a representative one that follows a self-supervised framework to achieve better effects than traditional DR algorithms \cite{hinton2006reducing}. A variety of embedding networks of different principles \cite{he2020momentum, chen2020simple} gradually arise for different data types \cite{guo2019deep, chen2020graph}. Literature reviews include many of the latest techniques \cite{li2018survey, abukmeil2021survey}. Most existing techniques, however, target unlabeled data, setting a reconstruction loss to make predicted samples gradually approach original ones, inapplicable in our scenario. Our approach needs to handle data features and user-specified labels jointly. The scenario is similar to those of supervised embedding techniques \cite{le2018supervised, li2018survey}. We analyze the differences between our network and these techniques in Section 4.

The importance of data projection for exploratory data analysis receives much attention in visualization \cite{wenskovitch2017towards, nonato2018multidimensional, sacha2016visual}. Researchers have proposed many strategies to improve pattern significance in data projections. Many works design new layouts or glyphs to generate visual summaries of adjacent points \cite{kammer2020glyphboard, van2015reducing, liao2017cluster}. However, the projection may not contain significant clusters, making determining grouping boundaries difficult. Subspace exploration is another feasible strategy. Xia et al. \cite{xia2017ldsscanner} proposed descriptors to construct the subspace to reveal low-dimensional structures unobservable in the original dimensional space. Many approaches allow the user to construct dimensions of the subspace interactively \cite{kim2015interaxis, gleicher2013explainers, li2021semanticaxis}. Each subspace dimension is the linear combination of many original dimensions. Thus, patterns found in the subspaces may not be easy to interpret. 

Many works are to explain patterns associated with specific visual structures in projections. Tian et al. \cite{tian2021using} proposed an approach to analyzing how specific patterns are distributed in projections. However, the approach targets specific patterns. Fujiwara et al. \cite{fujiwara2019supporting} proposed a contrastive learning-based algorithm, named ccPCA, to reveal which dimensions contribute more to the formation of a selected cluster. Many quantitative metrics, such as t-scores \cite{marcilio2021contrastive} and Shapley values \cite{marcilio2021explaining}, have also been used in this aspect. Faust et al. \cite{faust2018dimreader} explained how attribute value changes affect non-linear dimensionality reduction projections. Sohns et al. \cite{sohns2021attribute} utilized non-convex contours to highlight how a selected cluster varies under different attributes. However, these works require users to select candidate patterns, which can be difficult when projections do not contain clear visual structures.

\subsection{Knowledge-assisted Visual Analytics}
Chen et al. \cite{chen2008data} defined the data, information, and knowledge in visualization based on the DIKW pyramid \cite{rowley2007wisdom}. Besides, they proposed the concept of knowledge-assisted visualization that transfers knowledge in the human brain (perceptual and cognitive space) into control parameters of visualizations (computational space) through interactions. The two types of knowledge are tacit and explicit, proposed by Nonaka and Takeuchi \cite{nonaka2007knowledge}. Specifically, tacit knowledge is personal and exists in human brains only, while explicit knowledge is concrete and can be stored in a database. Most subsequent works utilize these concepts to model knowledge-assisted visual analytics \cite{van2005value, wang2009defining}. 

Incorporating knowledge into visual analytics has been on the visualization agenda \cite{thomas2005illuminating, keim2010mastering, thomas2006visual, chen2005top}. Many conceptual models give high-level blueprints. Stoiber et al. \cite{stoiber2022perspectives} classified these models as (1) descriptive and (2) mathematical. For (1), the knowledge generation model proposed by Sacha et al. \cite{sacha2014knowledge} is representative. Subsequent works extend the model to focus on specific analysis objects, such as objective and plan \cite{rind2016task} and uncertainty \cite{sacha2015role}. For (2), Van Wijk \cite{van2005value} proposed a model that describes the cost and gains of integrating knowledge into visual analytics. Many works have refined it \cite{green2009building, wang2009defining}. Federico et al. \cite{federico2017role} proposed the KAVA model that distinguishes tacit and explicit knowledge in Van Wijk’s model. Many works have applied the KAVA model in actual systems \cite{lohfink2021knowledge} and extended it to support new interactions \cite{ceneda2016characterizing}. The two types of models are the theoretical basis of our approach.

There is a wide variety of knowledge to implement the above models. Specifically, the knowledge can be standard procedures \cite{federico2015gnaeus} and linguistic rules \cite{nie2020knowledge} collected from domain literature, relationships between samples \cite{das2020geono, pister2020integrating, wall2017podium} specified by users, constraints distilled from expert experiences \cite{moritz2018formalizing}, numeric features calculated based on pre-collected samples \cite{wagner2017knowledge, wagner2018kavagait}, etc. Besides, many recent works attempt to utilize knowledge involved in off-the-shelf digital resources, such as ontology \cite{sobral2020ontology, lohfink2021knowledge}, corpus \cite{yang2020interactive}, knowledge graphs \cite{cashman2020cava, li2021kg4vis}, pre-trained models (e.g., knowledge distillation) \cite{wang2019deepvid}, etc. There have been literature reviews on techniques for integrating human knowledge into machine learning models \cite{deng2020integrating, von2019informed}, and many of them are also applicable in visualization. Along the line, knowledge in this paper derives from analysts' holistic understanding of the data (tacit knowledge), finally externalized as records labels (explicit knowledge).

How to introduce knowledge into visualization systems is another critical issue. Directly visualizing knowledge is a straightforward way. Federico et al. \cite{federico2015gnaeus} encoded clinical plans and actions acquired by experts as a tree to recommend diagnostic procedures. Wagner et al. \cite{wagner2018kavagait} visualized various gait patterns (knowledge) as references for identifying new abnormal patterns. A recent trend is introducing knowledge into the analysis models to break their performance bottleneck, yielding more desirable results. Relevant research involves many classic analysis tasks, such as labeling \cite{stewart2017label}, clustering \cite{yang2020interactive, pister2020integrating, das2020geono}, topic extraction \cite{kim2020architext, choo2013utopian}, ranking \cite{wall2017podium}, etc.  Our approach utilizes a similar strategy, i.e., incorporating human knowledge into the embedding network by adding a classification loss.

There are already works that utilize human knowledge to optimize loss \cite{diligenti2017integrating, clough2019topological} to lead the model to converge toward a more desirable direction. These works inspire us greatly. Unlike their focus on specific domains, our approach is domain-agnostic.

\section{Basic Idea}
\label{Basic Idea}
We introduce the idea of incorporating knowledge into embeddings and analyze its effects from the following aspects.

\subsection{Knowledge Definition}
Knowledge in this paper refers to the analyst's holistic understanding of the target dataset. When analyzing a dataset, the analyst can often judge the existence of specific patterns based on certain attributes. Using the Covid-19 dataset (see Figure~\ref{fig6}) as an example, the analyst may know that countries with different levels of Median Age or Per Capital GDP or on different continents may have different epidemic situations. The underlying rationales are (1) people in the same age range should have similar virus susceptibility, (2) the economic level determines how many resources a country can devote to controlling the epidemic spread, and (3) geographically proximate objects tend to be more related  \cite{tobler1970computer}. This situation is common in data analysis. Many datasets have attributes that lead analysts to think of specific patterns. Other common examples include gender (e.g., males and females have different preferences), week (e.g., weekdays and weekends have different criminal patterns), location (e.g., different regions have different climatic conditions), etc. These ideas (knowledge) derive from analysts' experiences and appear in the human mind spontaneously without complex calculations and analysis, namely the so-called tacit knowledge \cite{wang2009defining, federico2017role}.

The above knowledge definition has two characteristics:

First, knowledge formation involves dividing samples into groups according to their values on an attribute. For example, we put countries with different values on the three descriptive attributes into respective groups in the above Covid-19 cases. Below we discuss how to externalize tacit knowledge and incorporate it into an embedding network according to this characteristic (Sections 3.2 and 3.3).

Second, knowledge is holistic and irrelevant to individuals, indicating that counterexamples inevitably exist. For example, the analysts should know that many countries on the same continent have different epidemic situations since a continent spans a large area, whose countries often develop unevenly. We discuss how knowledge, especially counterexamples, affects data projections (Section 3.4).

\subsection{Knowledge Externalization}
\begin{figure}[htbp]
 \centering 
 \includegraphics[width=\linewidth]{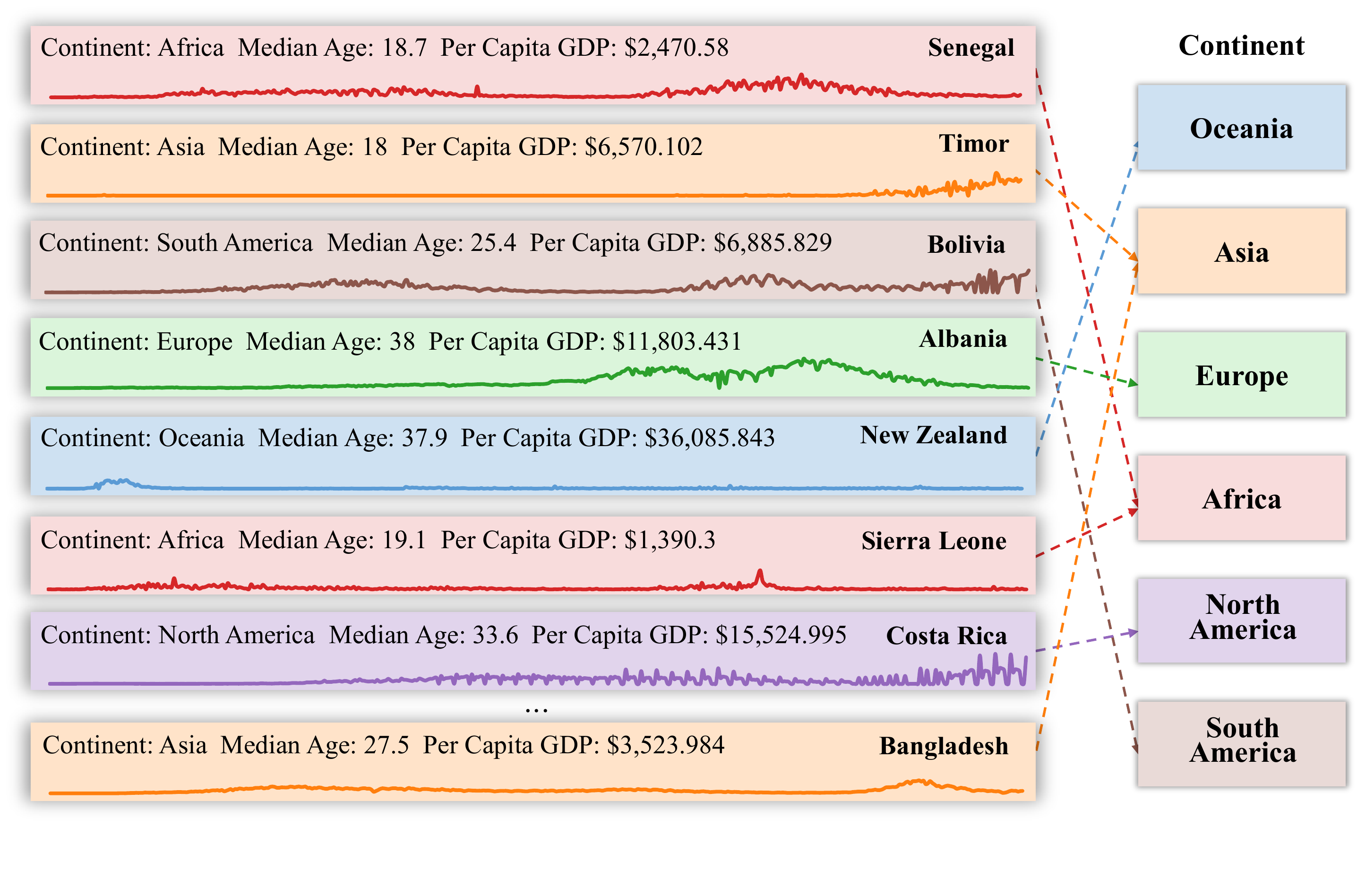}
  \put(-160,5){\small $(a)$}
  \put(-25,5){\small $(b)$}
    \vspace{-0.18in}
 \caption{Example of knowledge externalization on the Covid-19 dataset\cite{covid_19}. (a) Eight samples (countries). In addition to a sequence of monthly confirmed cases, each country has three attributes, i.e.,  Continent, Median Age, and Per Capital GDP, showing the country's statistical information during the period. (b) We assign countries of the same continent into a group (six groups in total), considering geographical proximate countries have similar epidemic situations.}
\label{fig2}
\end{figure}

Knowledge externalization converts analysts’ holistic understanding (tacit knowledge) into sample labels (explicit knowledge). We achieve it through (1) discretizing the value range of an attribute into intervals and (2) assigning samples with similar attribute values (falling in the same interval) into a group. Figure \ref{fig2} shows an example of knowledge externalization on the Covid-19 dataset\cite{covid_19}. The method is suitable for both numeric and categorical attributes. For a categorical attribute, each discretized attribute value corresponds to a group. As for a numerical attribute, we should first discretize its continuous value range by setting a binning resolution. A feasible solution is segmenting the value range according to statistical conventions or natural cycles.

\subsection{Knowledge Incorporation}

The embedding network should take the label and the original multi-dimensional feature as input. We achieve this by jointly using a reconstruction loss and a classification loss, as follows:

\textbf{Reconstruction loss}, denoted as $\ell_{r}$, is the basis for implementing the self-supervised framework of embedding networks. Let $x$ be a sample and $h$ be its embeddings. A self-supervised training gradually reduces the difference between $x$ and another reconstructed sample $x'$ output by a function with $h$ as the input. We can write $\ell_{r}$ as: 
\begin{equation}
\ell_{r}(x,h)=\Delta (x,f(h;\theta_{r})){}
\end{equation}
, where $f(h;\theta_{r})$ is the function for predicting $x'$, and $\theta_{r}$ represents the function parameters.

\textbf{Classification loss}, denoted as $\ell_{c}$, exists in supervised classifiers. Since knowledge externalization assigns each sample a label (group index), we can use $\ell_{c}$ in the embedding network to encode sample classes. Let $y$ be the class of $x$, which the analyst specifies. The training gradually reduces the difference between $y$ and $y'$ predicted by a function with $h$ as the input. We write $\ell_{c}$ as:
\begin{equation}
\ell_{c}(y,h)=\Delta (y,g(h;\theta_{c})){}
\end{equation}
, in which $g(h;\theta_{c})$ is the function for predicting $y'$, and $\theta_{c}$ represents the function parameters. 

We jointly use $\ell_{c}$ and $\ell_{r}$ as the loss of the embedding network, i.e.,
\begin{equation}
\ell=\alpha\ell_{c}+(1-\alpha)\ell_{r}{}
\end{equation}
, where $\alpha\in[0,1]$ is a parameter to adjust percentages of $\ell_{r}$ and $\ell_{c}$ in $\ell$. 
We discuss how $\alpha$ affects the projection effects in Section \ref{Data Embedding}.

\subsection{Effectiveness Analysis}
\label{Effectiveness Analysis}

\begin{figure}[htbp]
 \centering 
 \includegraphics[width=\linewidth]{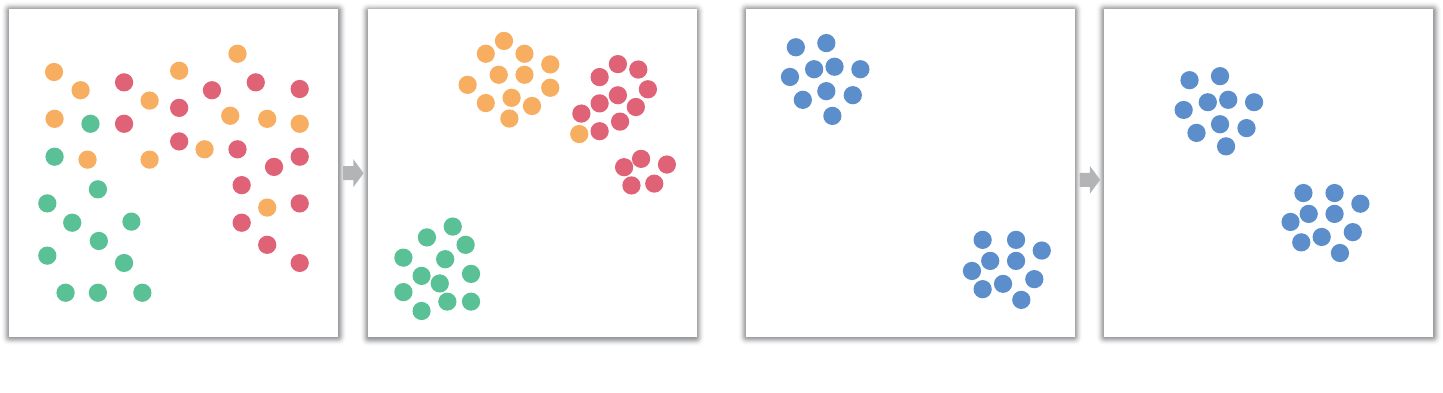}
     \put(-248,59){\small $a1$}
    \put(-186,59){\small $a2$}
    \put(-120,59){\small $b1$}
    \put(-57,59){\small $b2$}
   \put(-195,0){\small $(a)$}
  \put(-65,0){\small $(b)$}
 \vspace{-0.1in}
 \caption{Effects of incorporating knowledge into embeddings. (a) Samples have the same class and similar data features move close to form compact and class-consistent clusters. Visual structures, e.g., clusters and outliers, become more significant. (b) Assigning samples with different data features to the same group cannot form compact clusters.}
\label{fig3}
\end{figure}

We analyze why jointly using $\ell_{r}$ and $\ell_{c}$ can enhance the pattern significance and interpretability of the projection. Suppose there is a batch of samples, and we generated their embedding projection using our embedding network (Section \ref{Embedding Network}), as in Figure \ref{fig3}(a1). The embedding network initially only integrates $\ell_{r}$. Sample proximity reflects their data feature similarity. In other words, adjacent samples in the projection have higher data feature similarity. The color of each sample encodes its value on an attribute. We found that the red and orange samples overlap severely, relatively far from green. The finding indicates that the data feature similarity between red and orange samples is higher than between them and the green samples. Unfortunately, the projection does not contain any visual structure. Therefore, the analyst likely misses patterns when continuously changing the color encoding to get a snapshot of the corresponding data distribution.

We externalize knowledge by assigning samples of each color into a group and then regenerate the embeddings, as in Figure \ref{fig3}(a2). The proximity between samples thus reflects their similarity in data features and classes. In other words, samples of the same class move closer while samples of different classes move away from each other. $\ell_{c}$ acts as an invisible force to ensure that only samples of the same class with similar data features can form a compact cluster. 

The data projection thus has the following possible changes: (1) samples of the same class cluster together (see green and yellow clusters in Figure \ref{fig3}), indicating that these samples have similar data features; (2) samples of the same class form several small clusters (see two red clusters in Figure \ref{fig3}(a2)), indicating the existence of refined subclasses; (3) a few samples move far away from the majority of their class (see the yellow sample in the red cluster in Figure \ref{fig3}(a2)), indicating their distinctive data feature. These changes make the visual structures (clusters and outliers) clear and distinguishable, enabling better observing and inferring patterns according to their pairwise distances. For example, in Figure \ref{fig3}(a2), the red and yellow clusters are close to each other and relatively far from the green one, which is consistent with the pattern presented by the original projection (see Figure \ref{fig3}(a1)).

Since the dataset may contain counterexamples, incorrectly assigning samples with different data characteristics to a group is possible. Figure \ref{fig3}(b1) shows this situation. Analysts incorrectly assign samples of the two clusters to the same group. The knowledge thus pulls these samples with different data features closer, as in Figure \ref{fig3}(b2). However, their different data features prevent the formation of compact clusters, making the influence neglectable. 

A case on a synthetic dataset shows consistent results with the above analysis (see Section \ref{Synthetic Dataset}).

\section{Embedding Network}
\label{Embedding Network}

\begin{figure}[htbp]
 \centering 
 \includegraphics[width=\linewidth]{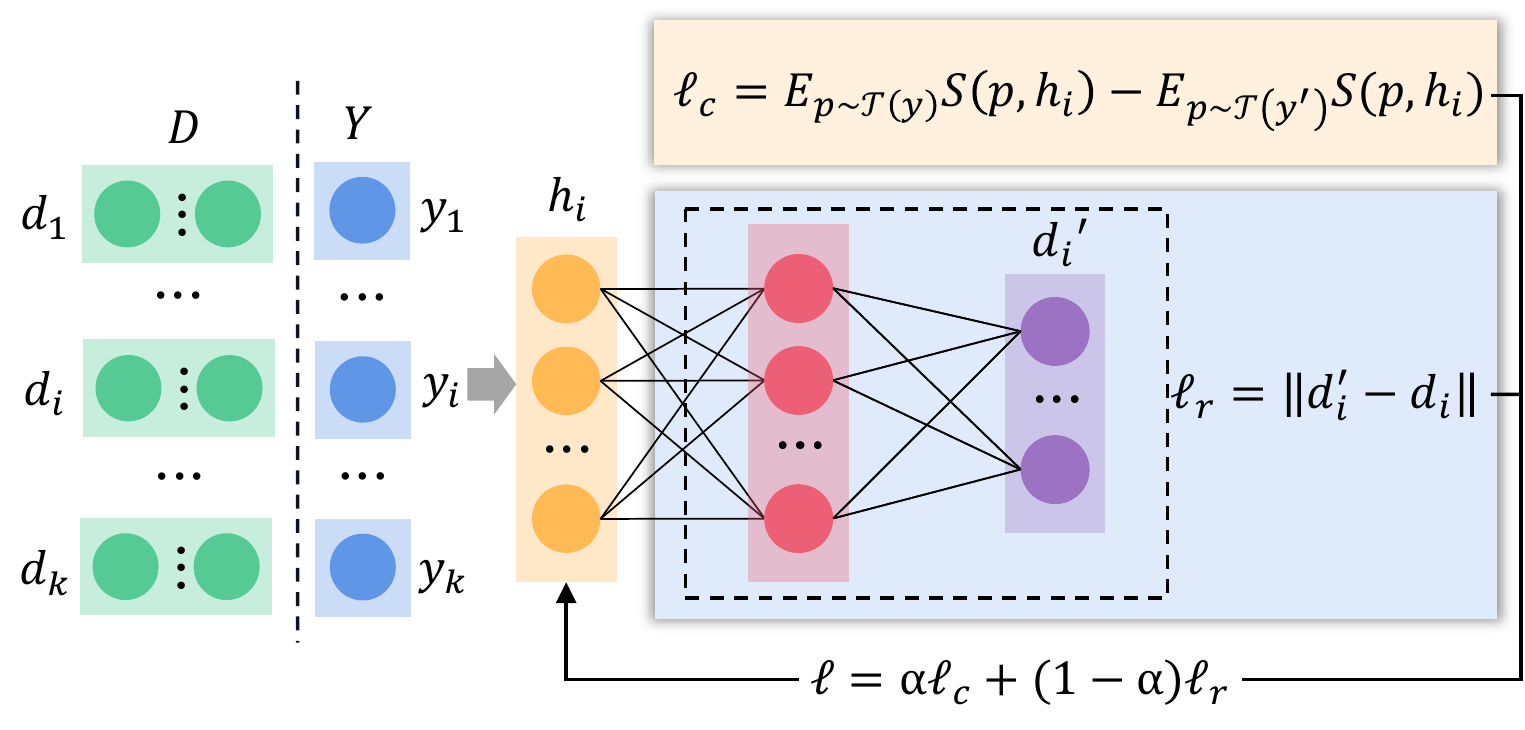}
 \vspace{-0.3in}
 \caption{Embedding network that jointly uses $\ell_{r}$ (blue background) and $\ell_{c}$ (orange background) to assign each sample an embedding. Each sample has a multi-dimensional feature (small green circles) and an analyst-specified label (small blue circles). }
\vspace{-0.1in}
\label{fig4}
\end{figure}

We propose an embedding network to implement the idea proposed in Section \ref{Basic Idea}, as in Figure \ref{fig4}. The network is a fully connected neural network with a single hidden layer (in red). Let $D$ be a dataset and $d_{i}$ be a sample of $D$ and $h_{i}$ be the embedding of $d_{i}$. The network takes $h_{i}$ as the input (in orange) to predict a reconstruction of $d_{i}$, denoted as $d_{i}'$. We initialize $h_{i}$ randomly at the beginning of the training. The network then updates $h_{i}$ by gradually decreasing the difference between $d_{i}'$ and $d_{i}$ during the training. We can define the reconstruction loss as: 

\begin{equation}
\ell_{r}=\left \| d_{i}'-d_{i} \right \|{}
\end{equation}

We then need to predict the label of $d_{i}$, denoted as $y_{i}'$, and calculate the difference between $y_{i}$ and $y_{i}'$ as $\ell_{c}$. Let $Y$ be the set of analyst-specified labels and $\mathcal{T}(y)$ be the sample group of label $y\in Y$ (consisting of samples with label $y$). We can get the similarity between $h_{i}$ and $\mathcal{T}(y)$ by calculating the average of similarities between $h_{i}$ and the embeddings of samples of $\mathcal{T}(y)$. We respectively calculate the similarity between $h_{i}$ and each sample group using the same method and choose the label of the group with the highest similarity to $h_{i}$ as the predicted label $y_{i}'$, as follows:

\begin{equation}
y_{i}'=argmax_{y \in Y}E_{p\in\mathcal{T}(y)}S(p,h_{i}){}
\end{equation}
, where $E$ represents the average of a group of similarities and $S()$ is an operator for calculating the similarity between two embeddings. 

Let $\mathcal{T}(y_{i}')$ be the group of samples of the predicted label $y_{i}'$, and $\mathcal{T}(y_{i})$ be the group of samples of the actual label $y_{i}$. We define the classification loss $\ell_{c}$ as:

\begin{equation}
\ell_{c}=E_{p\in\mathcal{T}(y_{i}')}S(p,h_{i})-E_{p\in\mathcal{T}(y_{i})}S(p,h_{i}){}
\end{equation}
, where $E_{p\in\mathcal{T}(y_{i}')}S(p,h_{i})$ and $E_{p\in\mathcal{T}(y_{i})}S(p,h_{i})$ be the average similarity between $h_{i}$ and the $\mathcal{T}(y_{i})$ and $\mathcal{T}(y_{i}')$, respectively.

Having obtained $\ell_{r}$ and $\ell_{c}$, we calculate the embedding loss $\ell$ using Equation 3. The network uses $\ell$ to update the embedding $h_{i}$ with gradient descent, as 
\begin{equation}
h_{i} \leftarrow h_{i}-\eta\frac{1}{k}
\begin{matrix} 
\sum_{i=1}^k\partial\ell
\end{matrix}/\partial h_{i}{}
\end{equation}
, where $\eta$ is the learning rate.


Our network differs from existing embedding techniques \cite{le2018supervised, li2018survey} in two aspects, as follows:

First, it integrates a customized $\ell_{c}$. Compared with widely-used cross-entropy, $\ell_{c}$ tends to improve intra-class similarities and reduce inter-class similarities of the samples, making it easy to form visual structures in the projection. The higher clustering accuracy than SAE (using cross-entropy) proves this point (see Section 8.1).

Second, it conducts network training and sample embedding simultaneously, unlike other techniques conducting the two steps separately (first training the network and then calculating embeddings using encoders). Our network does not have an encoder and updates embeddings (stored in a variable integrated as the input) in back-propagation, thus improving the embedding efficiency (see Section 8.2).

\section{Visualization System}
\label{Visualization System}
We integrate the embedding network into a visualization system to achieve a knowledge-based workflow, detailed below.

\subsection{Approach Overview}
\begin{figure}[htbp]
 \centering 
 \includegraphics[width=\linewidth]{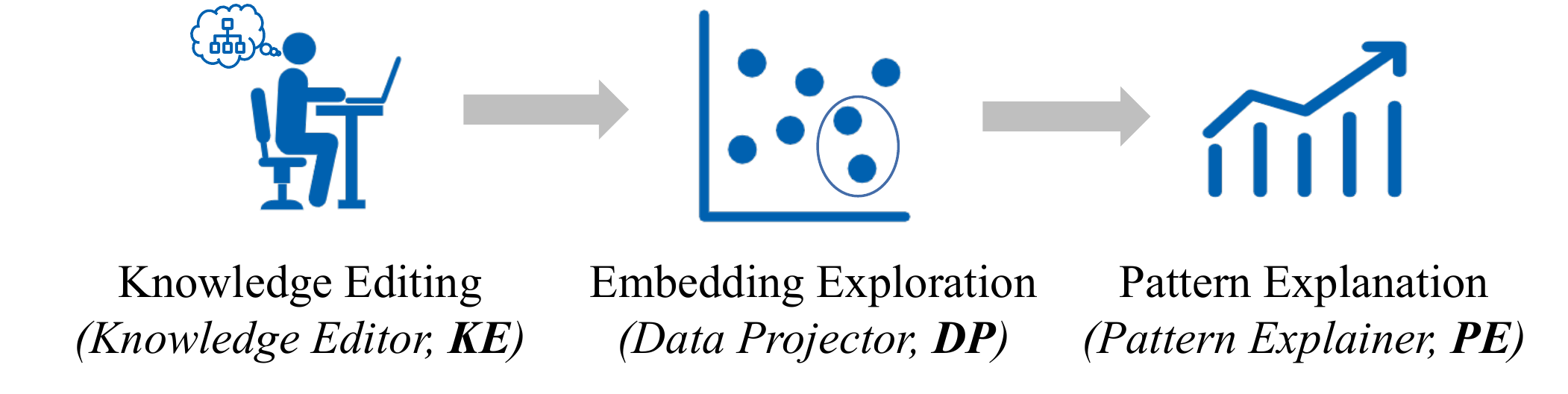}
  \vspace{-0.2in}
 \caption{The three steps and the corresponding visual components (in parentheses) in the exploratory workflow.}
\label{fig5}
\end{figure}

We give a three-step system workflow to achieve knowledge-based data analysis, as in Figure \ref{fig5}. Analysts (1) externalize tacit knowledge to generate embeddings, (2) select visual structures in the embedding projection, and (3) explain patterns of the visual structures.

We design a visualization system that integrates three components, namely Knowledge Editor (\textbf{KE}), Data Projector (\textbf{DP}), and Pattern Explainer (\textbf{PE}), to implement the three steps, as in Figure \ref{fig6}. We give design rationales for the three components as follows:

\textbf{KE} achieves sample grouping to externalize tacit knowledge. Specifically, the analyst can divide samples on one attribute to analyze high-level patterns (see examples in Section 3.1) or many attributes to explore fine-grained patterns. KE should provide visual cues to boost the formation of grouping ideas and a flexible grouping mechanism to achieve these ideas. (\textbf{R1, Grouping Flexibility})

\textbf{DP} shows the projection where the user selects visual structures for analysis. DP should support basic view operations (i.e., zooming, panning, lasso-selection, etc.) and rich interactions (e.g., sample highlighting and visual structure comparison). We should keep its interface concise, implement the interactions intuitively, and use as few controls as possible. (\textbf{R2, Operation Intuitiveness})

\textbf{PE} helps the analyst understand patterns associated with the selected visual structure. Specifically, it needs to explain how the visual structure forms (e.g., having similar epidemic situations in many months) and which group the selected visual structure refers to (e.g., countries in a continent). The information facilitates summarizing valuable conclusions. (\textbf{R3, Pattern Interpretability})

\subsection{Knowledge Editing}

KE follows a progressive strategy to achieve flexible grouping (R1). Below we introduce the visual design of KE together with its usage.

\begin{figure}[h]
 \centering 
 \includegraphics[width=\linewidth]{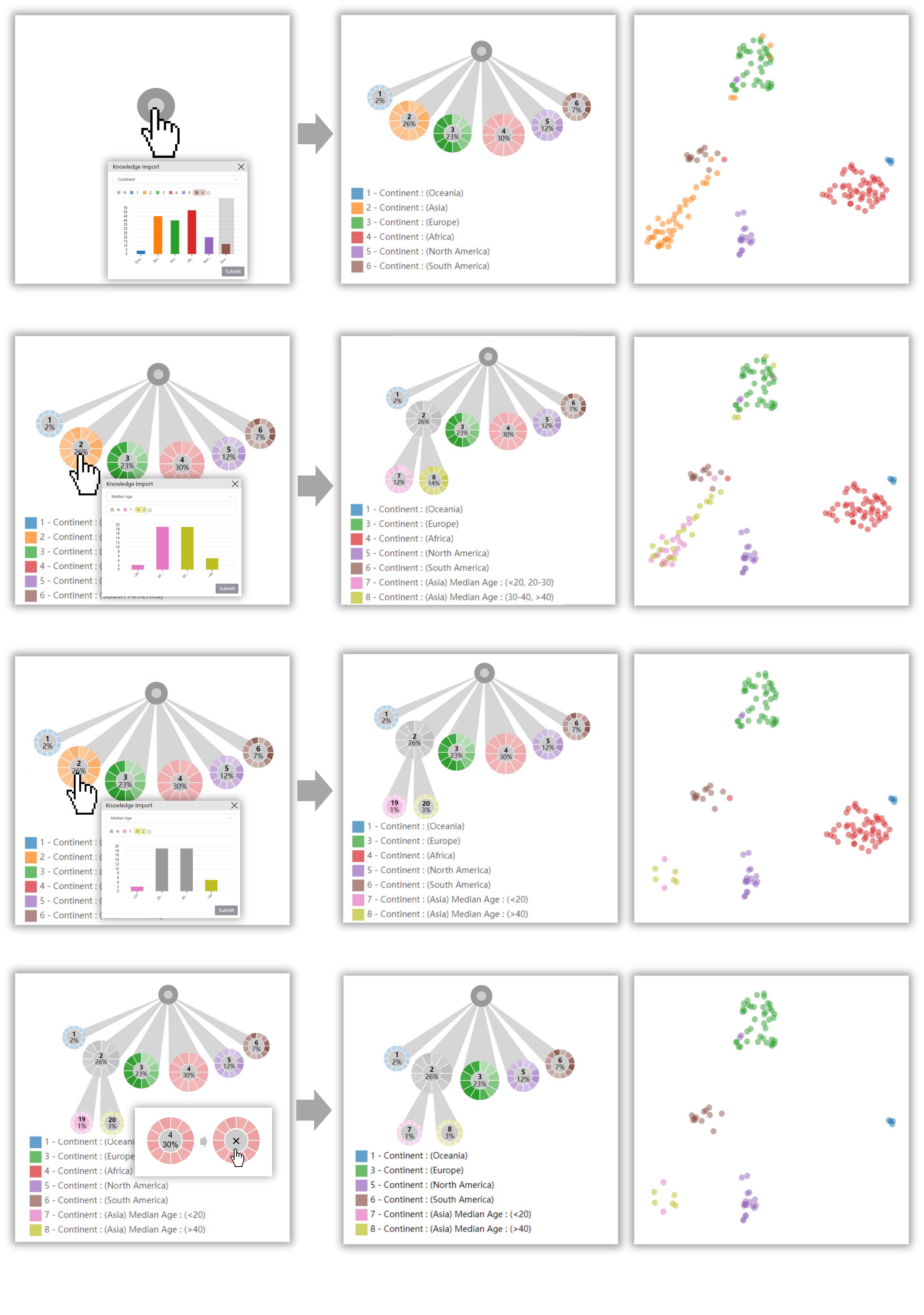}
  \put(-130,267){\small $(a)$}
 \put(-185,342){\small $a1$}
 \put(-95,342){\small $a2$}
 \put(-15,342){\small $a3$}
 \put(-130,180){\small $(b)$}
 \put(-185,253){\small $b1$}
 \put(-95,253){\small $b2$}
 \put(-15,253){\small $b3$}
 \put(-130,91){\small $(c)$}
 \put(-185,165){\small $c1$}
 \put(-95,165){\small $c2$}
 \put(-15,165){\small $c3$}
 \put(-130,3){\small $(d)$}
 \put(-185,78){\small $d1$}
 \put(-95,78){\small $d2$}
 \put(-15,78){\small $d3$}
  \vspace{-0.15in}
 \caption{Operations to realize KE's functions, including (a) class creation, (b) class refinement, (c) data filtering, and (d) class deletion. The left plot shows the analyst's operations in each subfigure, while the right two show the interface changes in KE and DP. }
 \vspace{-0.12in}
 \label{fig7}
 \end{figure}

\textbf{Class Creation.} We utilize the tree metaphor to implement KE. There is only the root in KE at the beginning of the data exploration, representing the whole dataset, as in Figure \ref{fig7}(a1). The analyst clicks on the root to activate a pop-up window, in which he (or she) can select an attribute to divide its value range into intervals for creating groups. A bar chart appears when the analyst selects an attribute to show the distribution of all samples on the attribute, as in Figure \ref{fig7}(a1). Each bar corresponds to a discretized value interval.

\begin{figure}[h]
 \centering 
 \includegraphics[width=\linewidth]{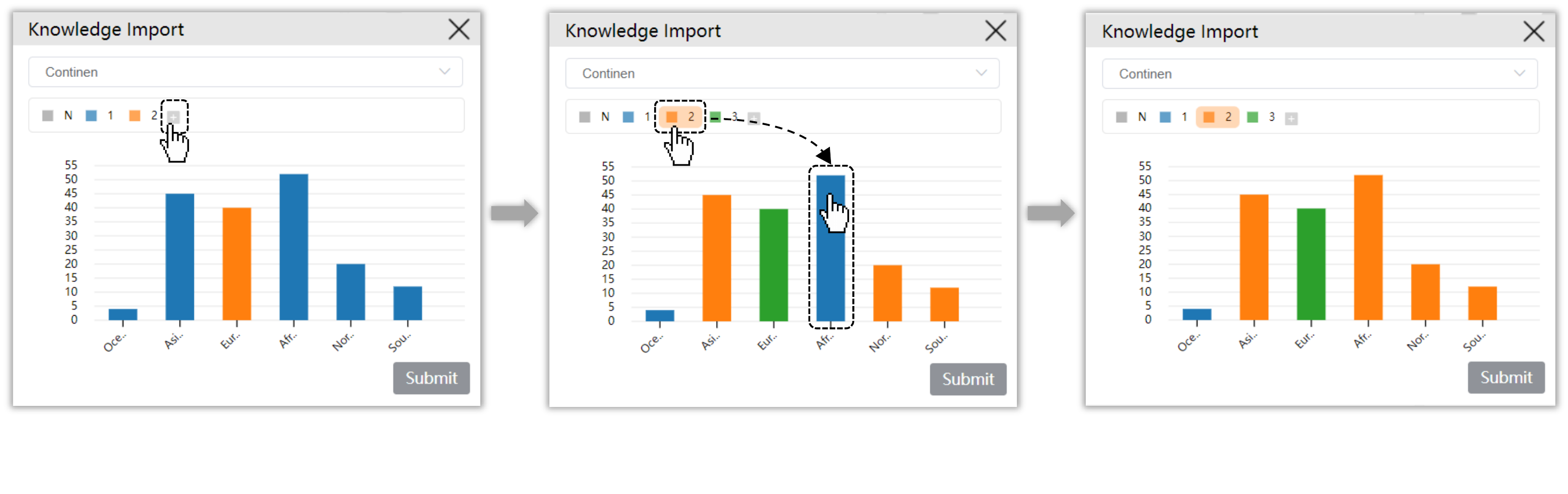}
    \put(-174,50){\small $(a)$}
    \put(-88,50){\small $(b)$}
 \vspace{-0.2in}
 \caption{Usage of the grouping control. (a) The analyst can click on a small grey rectangle to add a group. The control automatically clusters all bars (groups) into the specified number of groups. (b) The analyst then sequentially clicks on a colored rectangle (represents a group) and a bar to manually assign the bar to the group encoded by the rectangle.}
\label{fig14}
\end{figure}

KE utilizes clustering to simplify operations of creating groups. For this purpose, we generate a feature for each group (bar) by calculating the average of all group samples on each attribute for creating groups (e.g., the three attributes of the Covid-19 dataset). We encode each categorical feature as a one-hot vector and treat it as a multi-dimensional numeric feature, making the dimensionality of the group feature longer than the number of attributes for creating groups. When creating groups, the analyst sets the number of groups via a grouping control, as in Figure \ref{fig6}(a2). Then KE clusters all bars into groups via K-means (K equals the number of groups specified by the analyst), as in Figure \ref{fig14}(a). Clustering results serve only as a reference for the analyst, who can flexibly change the assignment of a bar by sequentially clicking on it and the colored rectangle of another group in the grouping control, as in Figure \ref{fig14}(b). Each cluster of bars corresponds to a class. We assign bars of the same class a unique color. Samples of a class (with attribute values within any interval of bars of the class) have the same color as the class, shown in DP (Section \ref{Data Embedding}), as in Figure \ref{fig7}(a3).

\begin{figure}[htbp]
 \centering 
 \includegraphics[width=0.9\columnwidth]{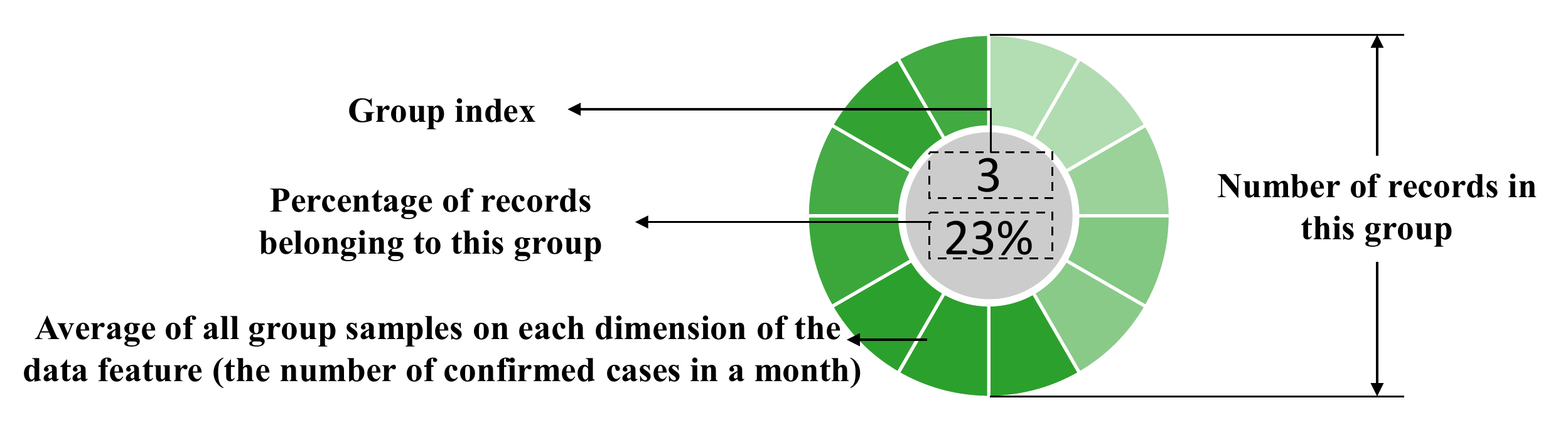}
 \vspace{-0.1in}
 \caption{Visual design of the tree node in the Knowledge Editor.}
  \vspace{-0.2in}
\label{fig15}
\end{figure}

KE encodes each class as a child node of the tree, as in Figure \ref{fig7}(a2). We implement each node as a pie chart. Each slice corresponds to a dimension of the data feature for embedding. The color of a slice encodes the average of all group samples on the dimension. For example, in Figure~\ref{fig15}, each slice shows the average confirmed cases of all group countries in a month (12 months in total). The darker the color, the higher the average on the dimension. The pie size encodes the number of contained samples. The pie center shows two numbers, i.e., (1) the class index of the pie and (2) the percentage of samples contained in the group to all samples. A legend at the bottom of KE shows each pie's color and value intervals.

\textbf{Class Refinement.} The analyst can view samples of a pie as a new dataset and create subclasses on it. He (or she) can click on the pie, and then the pop-up window used in creating classes appears again for creating subclasses, as in Figure \ref{fig7}(b1). Operations of creating subclasses are the same as those of creating classes.

KE encodes subclasses as child nodes of the selected pie, as in Figure \ref{fig7}(b2). Each subclass also has a unique color. After adding subclasses in the tree, the color of their father node becomes grey, indicating the class does not exist anymore, as in Figure \ref{fig7}(b2). The colors of samples belonging to the father class also change to colors of subclasses in DP (Section \ref{Data Embedding}), as in Figure \ref{fig7}(b3). The tree may have different levels of nodes by creating subclasses. Only colorful leaf nodes represent valid classes. For example, Figure \ref{fig7}(b2) contains seven valid classes.

\textbf{Data Filtering.} When using the pop-up window to create classes, the analyst can filter out bars to exclude the corresponding samples from the dataset. Figure \ref{fig7}(c1,c2) shows an example of creating two subclasses on parts of bars, leaving many grey bars that do not belong to any class (see the two grey bars in the pop-up window in Figure \ref{fig7}(c1)). Samples of these grey bars do not participate in the data analysis and disappear in DP also (Section \ref{Data Embedding}), as in Figure \ref{fig7}(c3).

\textbf{Class Deletion.} The analysts can delete a class. For this purpose, the analyst first clicks on the delete tab on the top of KE (see Figure \ref{fig6}(a1)) to enter class deletion mode, as in Figure \ref{fig7}(d1). In that mode, clicking on a pie can remove the corresponding class and the subclasses from the tree, as in Figure \ref{fig7}(d2). Samples of these deleted classes also disappear in DP, as in Figure \ref{fig7}(d3).

\subsection{Embedding Exploration}
\label{Data Embedding}
We implemented DP as a planar projection equipped with a set of interactions for facilitating the observation and selection of visual structures (R2). The embedding network runs at the backend and generates the embeddings after receiving the labels from KE. The analyst can adjust the percentage of $\ell_{c}$ in $\ell$ through a slider (see Figure \ref{fig6}(b1)) at the top of DP, as in Figure \ref{fig6}(b). We call the parameter Classification Loss Ratio (\textbf{CLR}, i.e., $\alpha$ in Equation 3). We set CLR to 0\% by default, and the analyst can interactively adjust the value during the data exploration. DP utilizes UMAP \cite{mcinnes2018umap} to project sample embeddings to form the projection. Other classic dimensionality reduction algorithms are also applicable. Each point in the projection represents a sample with the color indicating its class. DP supports fundamental view operations, such as zooming, panning, and lasso selection, facilitating the distribution observation and the sample selection.

\begin{figure}[htbp]
 \centering 
 \includegraphics[width=\linewidth]{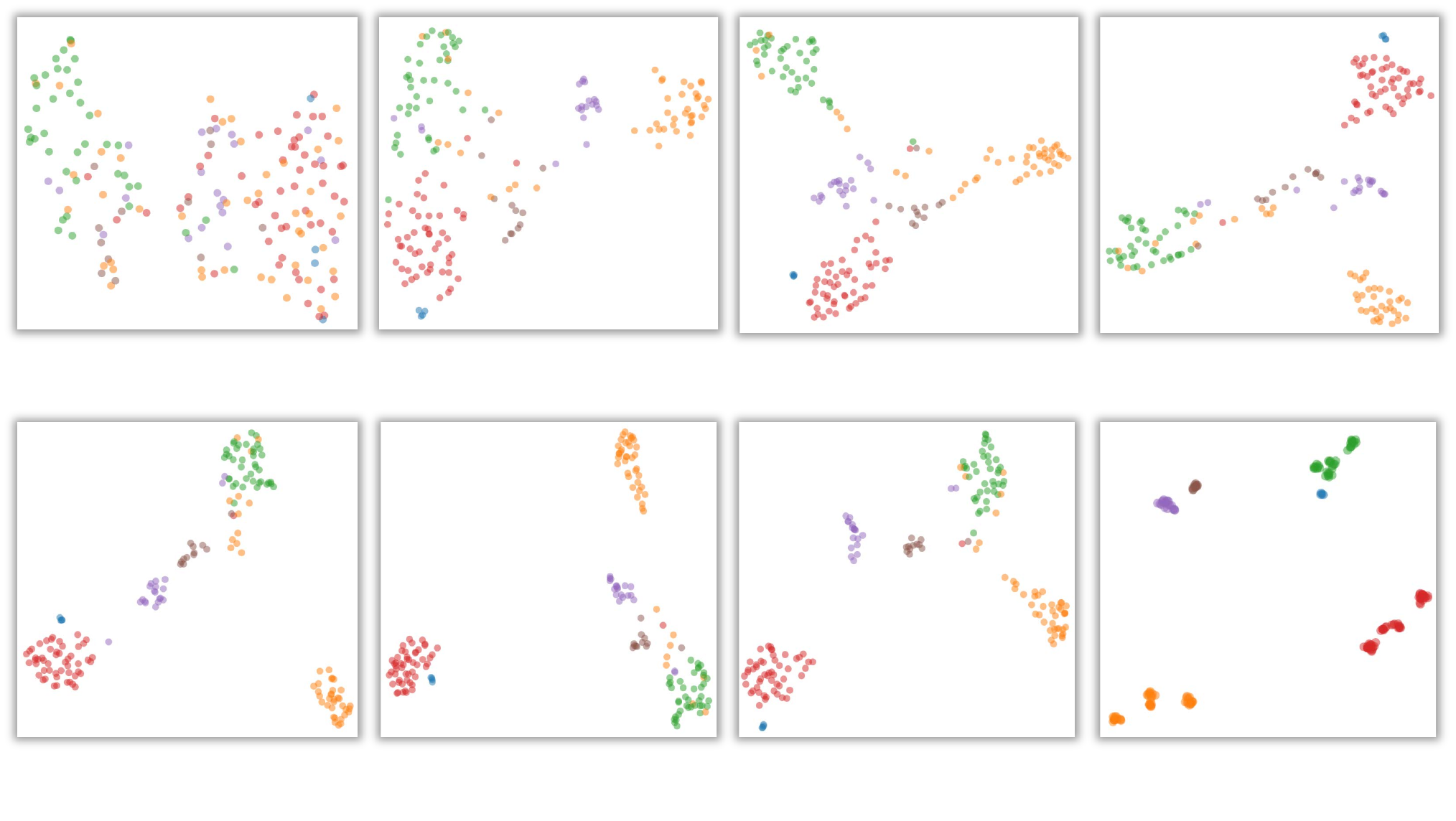}
     \put(-234,75){CLR=0}
    \put(-174,75){CLR=5\%}
    \put(-114,75){CLR=10\%}
    \put(-51,75){CLR=15\%}
    \put(-238,4){CLR=20\%}
    \put(-177,4){CLR=25\%}
    \put(-114,4){CLR=30\%}
    \put(-52,4){CLR=100\%}
  \vspace{-0.15in}
 \caption{Effects of different CLRs on the data distributions.}
 \vspace{-0.1in}
 \label{fig8}
\end{figure}

Figure \ref{fig8} shows the projection of the Covid-19 dataset\cite{covid_19} under different CLRs. We divide all countries into six groups according to their continents. Each color represents a continent. A significant phenomenon is that clusters gradually become clear as CLR increases. Therefore, it is more convenient to see relationships among groups by setting a large CLR. Two adjacent clusters should have a high similarity concerning the data features of their samples. 

It is unnecessary to find an optimal CLR during the data exploration. In contrast, we think different CLRs have their usages, even for two extreme values, i.e., 0\% and 100\%. $\ell_{c}$ does not work when setting CLR to 0\%. However, the projection can help analysts understand the original data distribution. In contrast, we can set CLR to 100\% to observe the classification information. Typically, the CLR should lie between the two extreme values. 

We provide two tips on how to set CLR. First, the analyst can set a small CLR (e.g., \textless10\%) to see which samples still form clusters, indicating strong patterns among them. These samples must have similar data features since a low CLR that makes $\ell_{c}$ play a minor role in forming clusters that can still pull them together. Second, the analyst can set a large CLR (e.g., \textgreater30\%) to identify outliers that may be very different from other samples of the same classes, as even a high CLR cannot make them close.

\subsection{Pattern Explanation}
The analyst can lasso-select two visual structures in DP (see those enclosed by blue and pink lassos in Figure \ref{fig6}(b)) and utilize PE to compare them. We calculate a SHAP value for each attribute by training a classifier to distinguish samples of the two visual structures. A high SHAP value (absolute value) indicates that the attribute is vital in separating the two visual structures. PE allows the analyst to select a single visual structure and treats the unselected samples as another visual structure to train the classifier. Thus, the SHAP values indicate how the selected samples differ from others.

PE shows a group of vertically-aligned factors. Each factor is an attribute of the dataset. From top to bottom, their SHAP values  \cite{lundberg2017unified} gradually decrease. PE supports two types of factors. We consider (1) each dimension of the embedded data feature as an Embedding Factor (\textbf{EF}) and (2) each value interval of an attribute for externalizing knowledge as a Classification Factor (\textbf{CF}) (R3). Using Figure \ref{fig2} as an example, the number of confirmed cases in a month is an EF, while a continent is a CF. EF and CF play different roles in explaining the formation of visual structure. An EF explains how the two visual structures form (Why). Meanwhile, a CF can tell analysts which groups the two visual structures refer to (Who). The analysts can switch the factor list to show either CFs or EFs by clicking the corresponding tabs on the top of the interface, as in Figure \ref{fig6}(c1).

\begin{figure}[htbp]
 \centering 
 \includegraphics[width=\linewidth]{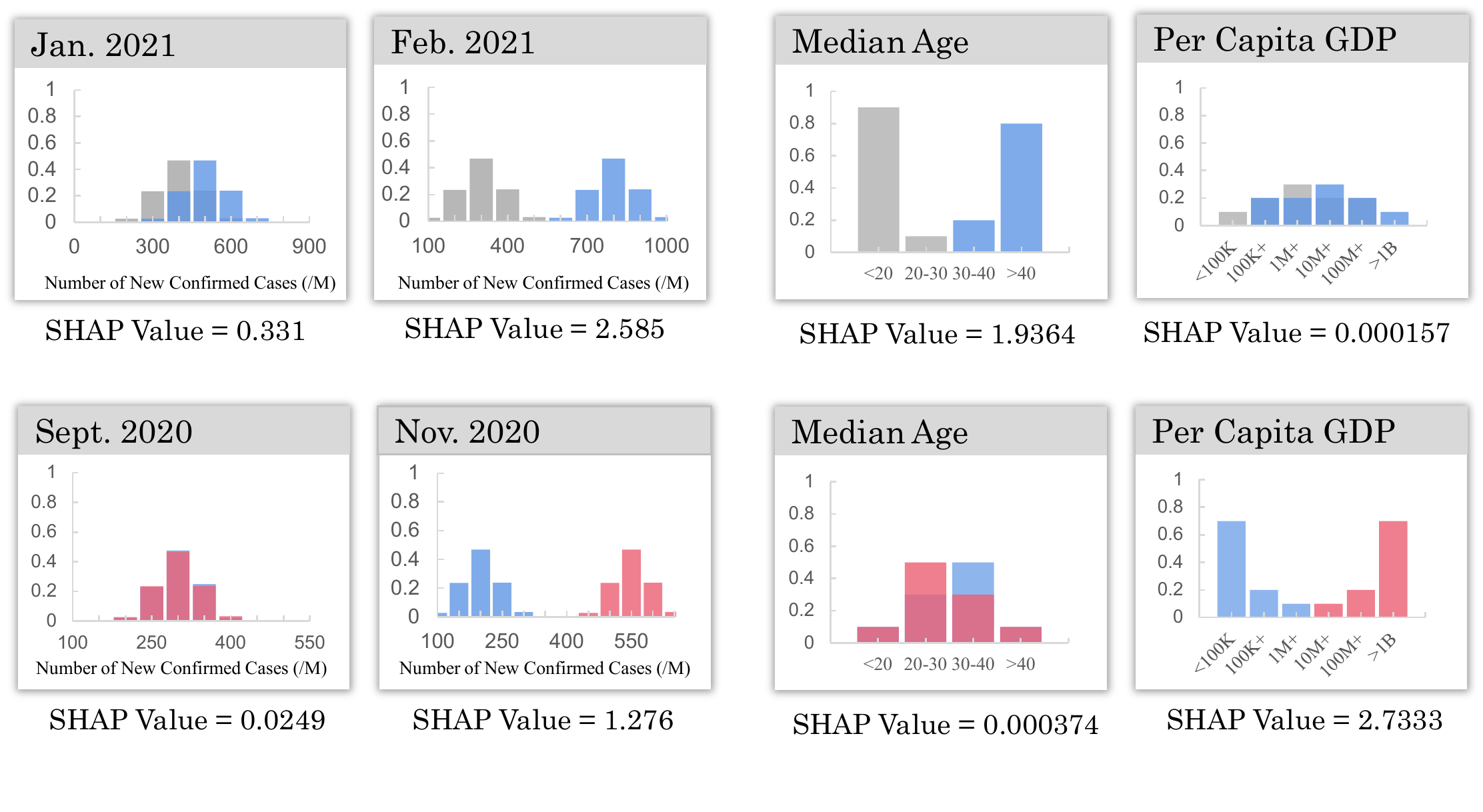}
    \put(-195,75){\small $(a)$}
    \put(-66,75){\small $(b)$}
    \put(-195,7){\small $(c)$}
    \put(-66,7){\small $(d)$}
    \vspace{-0.2in}
 \caption{Histograms for explaining the formation of the selected visual structures. The analyst can select one visual structure to compare its distributions (in blue) with unselected samples (in grey) on (a) EF or (b) CF. He (or she) can also select two visual structures to compare their distributions (in blue and pink respectively)  on (c) EF or (d) CF. In all subfigures, the lower the SHAP values, the more similar the two distributions, i.e., the factor contributes little to separate the two distributions.}
 \vspace{-0.2in}
 \label{fig9}
\end{figure}

The analyst can click on a factor to generate a histogram to show the distributions of the selected visual structures on the factor, as in Figure \ref{fig6}(c2). There are always two distributions in each histogram, one for the selected visual structure (in blue) and the other for the unselected samples (in grey) or the second selected visual structure (in pink). The histogram verifies the importance of a factor. The two distributions should separate in the histogram when the SHAP value is high and gradually overlap as the SHAP value decreases, as in Figure \ref{fig9}.


\subsection{Implementation}
The visualization system has a client-server architecture implemented using the Flask. The frontend interface is written using D3\cite{bostock2011d3}. The embedding network, trained using TensorFlow\cite{Abadi_TensorFlow_Large_scale_machine_2015}, runs at the backend, receiving data labels from the frontend via the FLask and sends back the embeddings to the frontend for visualization.

\section{Usage Scenario}
\label{Case Study}
We show the usage scenarios of our approach on a synthetic dataset and two real-world datasets. We set CLR to 20\% in all experiments.

\subsection{Synthetic Dataset}
\label{Synthetic Dataset}
\begin{figure}[htbp]
 \centering 
 \includegraphics[width=\linewidth]{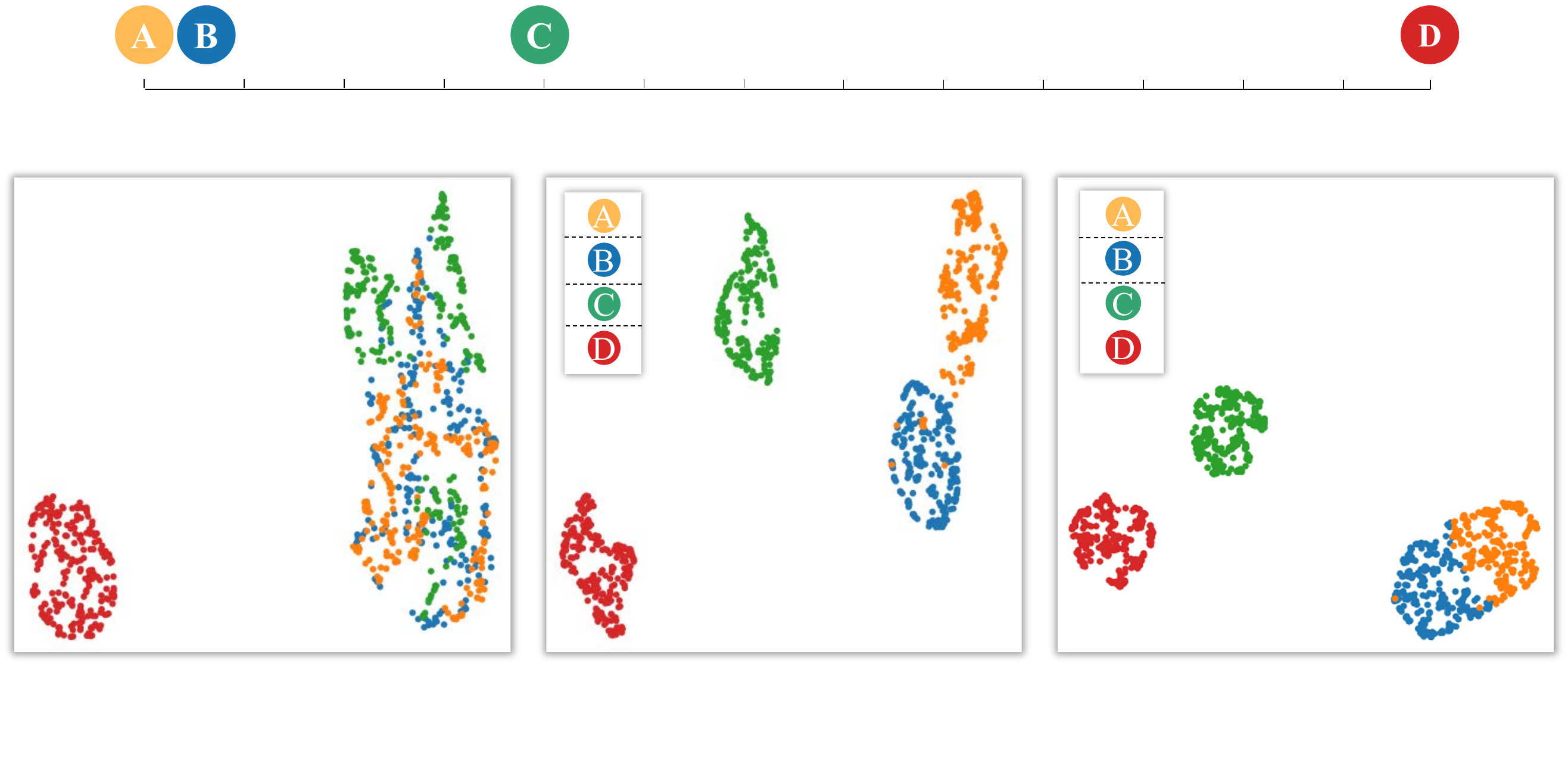}
     \put(-132,100){\small $(a)$}
    \put(-214,10){\small $(b)$}
    \put(-132,10){\small $(c)$}
    \put(-46,10){\small $(d)$}
    \vspace{-0.25in}
 \caption{Experiment results on a synthetic dataset. (a) Illustration of data feature similarity among the four groups. (b) Original data distribution generated using UMAP. (c) Projection created by correctly assigning different labels to the four groups of samples. (d) Projection created by incorrectly assigning the same labels to the samples of C and D.}
 \label{fig10}
\end{figure}

We utilized a synthetic five-dimensional dataset to illustrate the effectiveness of the general idea. We preset four groups (A, B, C, and D) whose samples have consistent and mutually different attribute values. Each group has 250 samples. We randomly generate five numbers within a user-specified value range as a sample, while samples of a group have the same value range. We make the value ranges of the four groups overlap. The more the value ranges of two groups overlap, the higher the data feature similarity they have. In general, B is the most similar to A (60\% value ranges overlap on each attribute), weakly similar to C (40\% value ranges overlap on each attribute) but differs from D entirely (without overlapping value ranges), as in Figure \ref{fig10}(a). Figure \ref{fig10}(b) shows the original sample distribution generated using UMAP. We can see that samples of A, B, and C overlap but are separate from those of D. Analysts thus can hardly know the pattern that A and B are weakly similar to C with the projection.

We assigned samples of each group to a class and incorporated the four classes into the embedding network via $\ell_{c}$, as in Figure \ref{fig10}(c). The original large cluster (containing A, B, and C) splits into three small clusters of respective groups. Moreover, the proximity between two clusters reflects the overall similarity of the two groups. Analysts thus can easily derive the preset patterns.

We then recreated classes to conduct another experiment. We assigned samples of A and B to respective classes but merged C and D as a single class. The experiment helps in understanding how counterexamples affect the embedding results. Figure \ref{fig10}(d) shows the  projection. We found that C and D move closer but are still separate from each other due to their differences in attribute values, which is consistent with our analysis (Section \ref{Effectiveness Analysis}).

\subsection{Global COVID-19 Dataset}
Figure \ref{fig2} shows the data structure. We incorporate knowledge of three attributes (i.e., continent, median age, Per Capital GDP) and explore patterns within the embedding projections as follows:

\textbf{Continent.} We classified all countries into six classes according to their continents (Figure \ref{fig6}(a)) and generated the projection (Figure \ref{fig6}(b)) in the visualization system. Countries of each continent form a separate cluster. We then conducted pairwise cluster comparisons to understand epidemic situations of different continents. The analyst selected Africa and South America clusters in Figure \ref{fig6}(b). PE shows EFs (number of monthly confirmed cases) sorted according to their SHAP values from high to low. We selected the month with the highest SHAP values to generate a histogram(see Figure \ref{fig6}(c2)) and found that countries in Africa (in pink) have fewer confirmed cases than those in South America (in blue), as in Figure \ref{fig6}(c). We repeatedly compared other continents with the same methods, finally understanding the epidemic situations of different continents during that period. Oceania and Africa generally have the fewest confirmed cases, while the epidemics in Europe and South America are severe. Asian and North American countries vary in severity: most countries (in the big orange and purple clusters) have few confirmed cases, and a few (close to the green cluster) have similar severity to Europe.

\begin{figure}[htbp]
 \centering 
 \includegraphics[width=\linewidth]{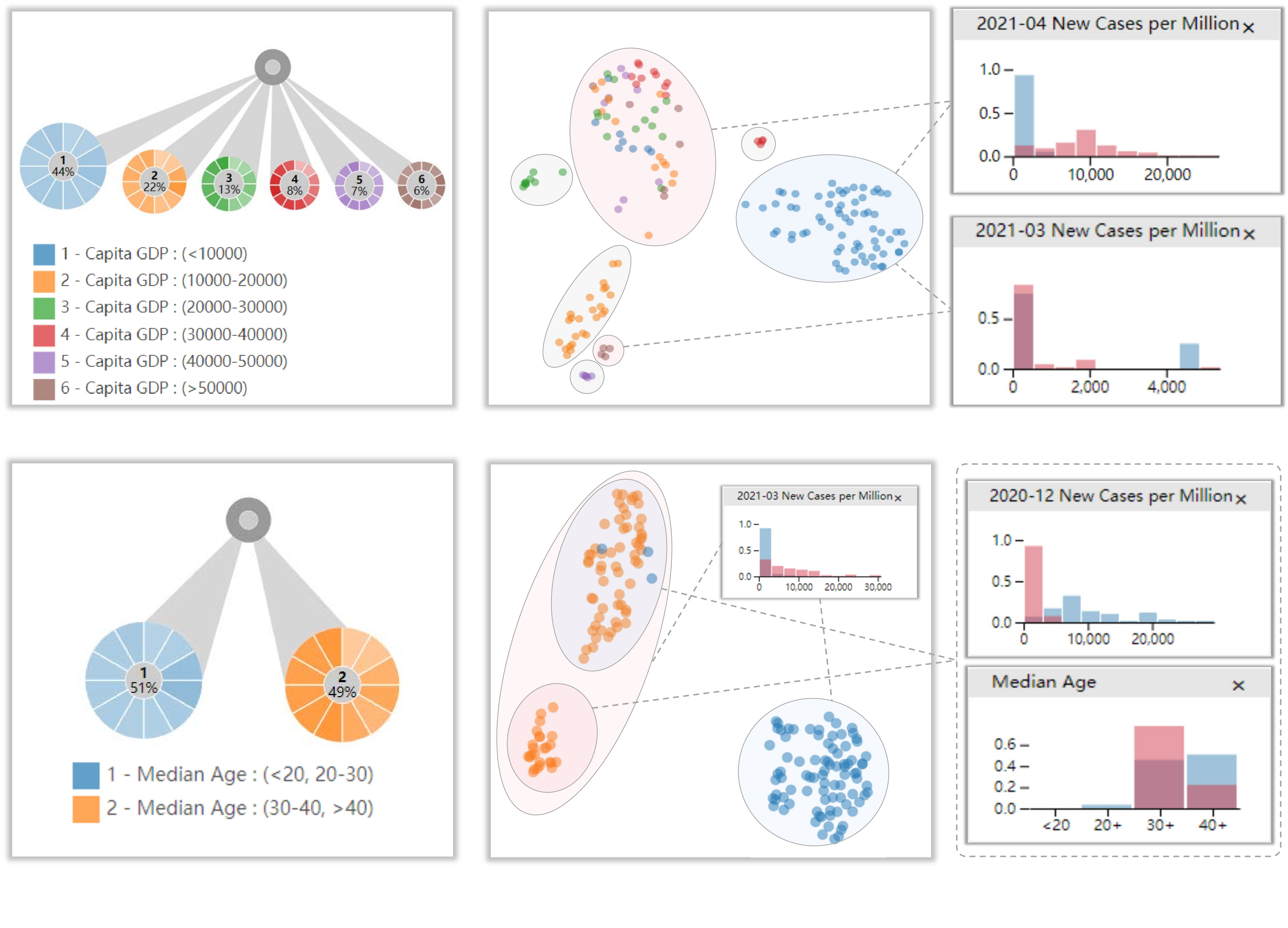}
    \put(-210,95){\small $(a)$}
    \put(-118,95){\small $(b)$}
    \put(-40,95){\small $(c)$}
    \put(-210,5){\small $(d)$}
    \put(-118,5){\small $(e)$}
    \put(-40,5){\small $(f)$}
    \put(-135,143){\small $b1$}
    \put(-130,112){\small $b2$}
    \put(-145,103){\small $b6$}
    \put(-145,123){\small $b5$}
    \put(-155,148){\small $b4$}
    \put(-110,155){\small $b7$}
    \put(-98,127){\small $b3$}
    \put(-10,166){\small $c1$}
    \put(-10,126){\small $c2$}
    \put(-13,74){\small $f1$}
    \put(-13,37){\small $f2$}
    \put(-150,50){\small $e1$}
    \put(-143,60){\small $e3$}
    \put(-143,39){\small $e4$}
    \put(-81,76){\small $e5$}
    \put(-106,29){\small $e2$}
    \vspace{-0.16in}
 \caption{Cases found by externalizing knowledge on (a-c) Per Capital GDP and (d-f) median age.}
 \label{fig11}
\end{figure}

\textbf{Per Capital GDP.} According to statistical conventions, we created six classes on Per Capital GDP, as in Figure \ref{fig11}(a). Figure \ref{fig11}(b) shows the projection in which we found a cluster consisting of countries with diverse Per Capital GDP levels (b1) and multiple color-consistent clusters (b2-b7). We selected these clusters pair by pair to compare their epidemic situations. We found (1) countries in b1 have more confirmed cases than other countries, and (2) countries in b2-b7, though with mutually different Per Capital GDP levels, have similar epidemic situations. Two histograms (Figure \ref{fig11}(c1, c2)) generated by the EFs with the highest SHAP value prove these findings by showing comparative results of b1-b3 and b2-b3. 

\textbf{Median Age.} We created two classes (low media age and high media age) with similar numbers of countries (Figure \ref{fig11}(d)) and incorporated the classification information into the network. We found that countries with low median age are mainly in a single cluster (in blue), while high median age countries form two adjacent clusters (in orange), as in Figure \ref{fig11}(e). We selected two separate orange clusters as a whole and compared them with the blue one. The EF with the highest SHAP value shows the result. Its histogram shows that countries with low median ages (in blue) have fewer confirmed cases, as in Figure \ref{fig11}(e5). We then selected the two orange clusters as individual groups and analyzed their differences. Although belonging to the high median age class, the two clusters still differ in a few aspects: countries in the upper cluster have more confirmed cases (EF) and higher median ages (CF) than countries in the lower cluster, as in Figure \ref{fig11}(f1, f2). These findings show the correlation between median age and the epidemic severity.

\begin{figure*}[htbp]
 \centering 
 \includegraphics[width=\linewidth]{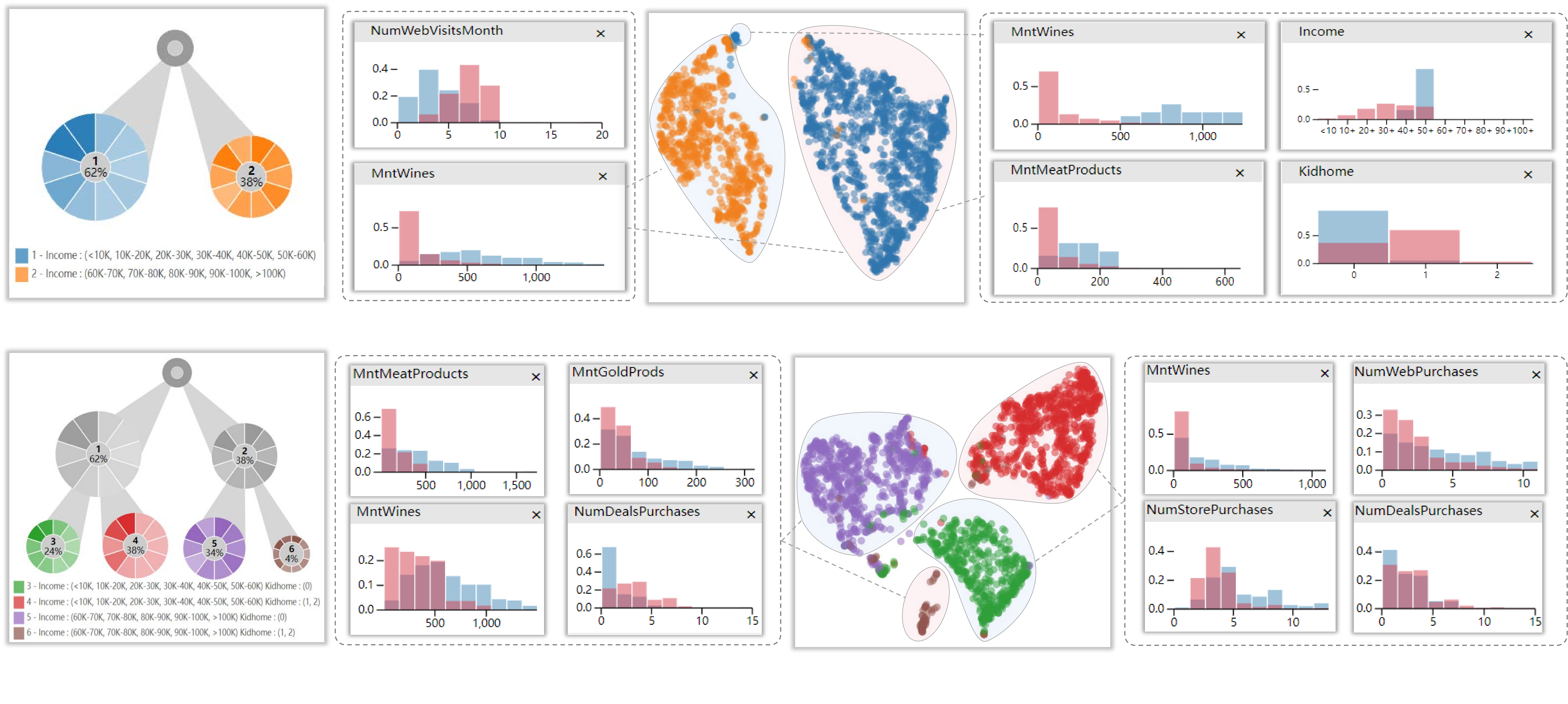}
     \put(-465,119){\small $(a)$}
    \put(-360,119){\small $(b)$}
    \put(-257,119){\small $(c)$}
    \put(-102,119){\small $(d)$}
    \put(-465,10){\small $(e)$}
    \put(-339,10){\small $(f)$}
    \put(-208,10){\small $(g)$}
    \put(-79,10){\small $(h)$}
    \put(-320,205){\small $b1$}
    \put(-320,158){\small $b2$}
    \put(-265,178){\small $c1$}
    \put(-230,205){\small $c2$}
    \put(-270,215){\small $c3$}
    \put(-109,205){\small $d1$}
    \put(-109,158){\small $d2$}
    \put(-16,205){\small $d3$}
    \put(-16,158){\small $d4$}
    \put(-346,97){\small $f1$}
    \put(-275,97){\small $f2$}
    \put(-346,51){\small $f3$}
    \put(-275,51){\small $f4$}
    \put(-86,97){\small $h1$}
    \put(-18,97){\small $h2$}
    \put(-86,51){\small $h3$}
    \put(-18,51){\small $h4$}
    \vspace{-0.23in}
 \caption{Cases found in customer behavior dataset. (a-d) Exploring how annual income affects purchase behaviors by setting two classes, i.e., high-income and low-income. (e-h) Adding subclasses to explore more fine patterns. The background color of each selected visual structure is the same as the colors of its bars in histograms.}
 \vspace{-0.18in}
 \label{fig12}
\end{figure*}
\subsection{Customer Behavior Data}

The dataset\cite{customer_behavior} consists of 2214 samples, each corresponding to a customer. Each sample contains two types of information, i.e., personal information (CFs) and purchase behavior (EFs) in marking campaigns. We picked out four attributes of personal information, i.e., (1) birth year, (2) education level, (3) number of kids, and (4) annual income. And the purchase behavior involves ten attributes, i.e., (1-2) number of online/offline purchases, (3) number of shopping website visits, (4) number of discounted purchases, (5-10) amount spent on wine/fruits/meat/fish/sweets/gold products. 

We divided customers into two groups according to their annual incomes, considering that high- and low-income people should have different purchasing behaviors, as in Figure \ref{fig12}(a). We imported the two classes into the embedding network with ten purchase-behavior attributes to generate embeddings. Figure \ref{fig12}(c) shows the projection in which the two classes form two clusters, confirming the different purchasing behaviors of their respective customers. We found that two EFs, i.e., MntWines and NumWebVisitMonth, have the highest SHAP values. Their histograms show that low-income customers visit shopping websites more (Figure \ref{fig12}(b1)) but buy fewer products than high-income customers (Figure \ref{fig12}(b2)).


We also found a small cluster of low annual income customers, as in Figure \ref{fig12}(c3). We compared this small cluster to the large one (Figure \ref{fig12}(c2)). 
We utilized PE to observe the distribution of the two EFs with the highest SHAP values, and we found that most customers in the small cluster buy more products (Figure \ref{fig12}(d1, d2)). We then selected the two CFs with the highest SHAP values (number of kids and annual income) and observed the distributions of the two clusters. We found that the customers in the small cluster have higher incomes (Figure \ref{fig12}(d3)) and do not have a kid (Figure \ref{fig12}(d4)).

The above findings inspire us to explore whether having kids affects purchase behaviors. We created two subclasses (i.e., with kids and without a kid) for each class created on annual income, forming a two-level tree with four classes, as in Figure \ref{fig12}(e). We conducted two comparative tasks, i.e., (1) low income with kids (in red) VS low income without kids (in green) and (2) high income with kids (in brown) VS high income without kids (in purple), as in Figure \ref{fig12}(g). We picked EFs with the highest SHAP values for the two tasks to generate histograms. We found that high-income and low-income customers spend less on wine (see Figure \ref{fig12}(f3) and Figure \ref{fig12}(h1)) and buy more discounted products (see Figure \ref{fig12}(f4) and Figure \ref{fig12}(h4)) after having kids. However, high-income customers buy fewer gold products and meat when having kids (Figure \ref{fig12}(f1, f2), while low-income customers reduce their purchases at both websites and stores (Figure \ref{fig12}(h2, h3)).  

\section{User Study}
\label{User Study}

We conducted a user study to understand how analysts use the visualization system in real scenarios.
\subsection{Experiment Setup}
We discuss the experiment designs from the following aspects:

\textbf{Participants.} We recruited 20 subjects (16 males and 4 females; aged 22-26, median 23). All subjects are graduate students majoring in machine learning and data mining from our school. They have experience in data analysis and knowledge of representation learning but are not involved in this project.

\textbf{Task.} The task for each participant is to find three country groups from the Covid-19 dataset and use PE to compare their epidemic situations. Specifically, most countries in each group should have the same values on a CF, and the participant needs to find the EFs (month) with the highest SHAP values in pairwise comparisons using PE.


\textbf{Experiment Method.} We chose a between-subject design to compare subjects' performance when using the system that allows and does not allow knowledge incorporation. We divided all subjects into two groups of equal size, GA and GB. GA subjects can use the system without any limitation, while CLR is always zero to block the knowledge incorporation when GB subjects use the system.

\textbf{Training.} There is a training session before the experiment. An administrator briefly introduced the basic idea of our approach and demonstrated how to use the system to find a pattern from the customer behavior dataset. Then subjects use the system to find a pattern themselves. An experimenter is responsible for answering questions from the subjects and ensuring all subjects are proficient in using the system.

\textbf{Process.} Each subject sequentially (1) creates groups, (2) adjusts CLR (GA only) to generate the projection, and (3) lasso-select groups (visual structures). He (or she) then (4) explains patterns, i.e., comparing two groups by generating histograms of the EF and the CF with the highest SHAP values, and (5) saves results, i.e., putting screenshots into a document and writing a simple annotation on the paper. The subject can regenerate the projection and perform the subsequent steps when he (or she) finds it impossible to select satisfactory visual structures. The subject can terminate the experiment when finding three groups. We set a screen recording software to record each subject's experimental process. There is also a questionnaire session after the experiment. Each subject needs to score their overall satisfaction with the system (1-10) and write subjective comments in the questionnaire.


\subsection{Experiment Results} 
\begin{figure}[htbp]
 \centering 
 \includegraphics[width=\linewidth]{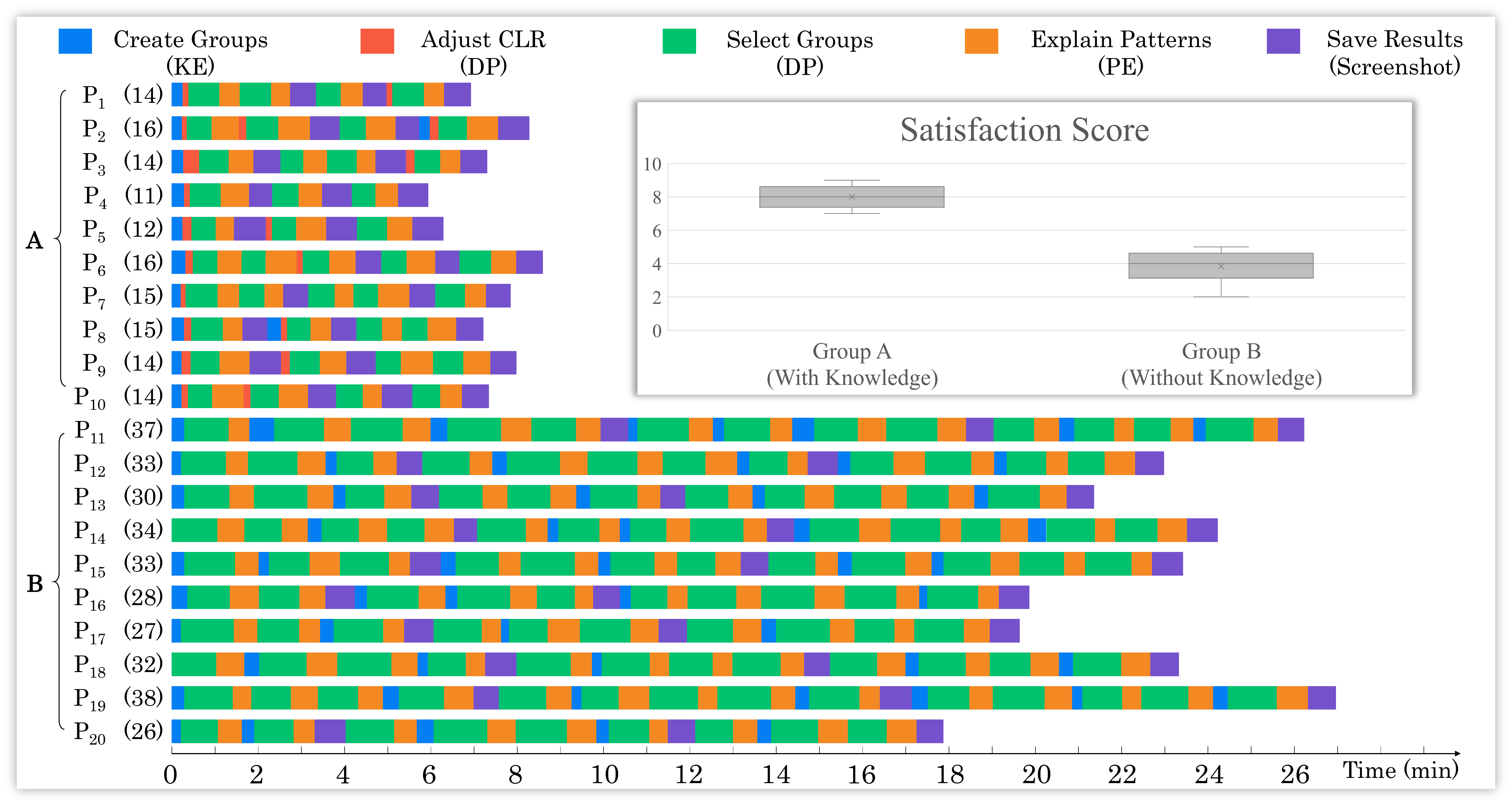}
 \put(-30,107){\small $(b)$}
 \put(-12,112){\small $(a)$}
 \vspace{-0.1in}
 \caption{User study results. (a) Operation logs of subjects in completing the preset task using the visualization system. Color encodes operation type. (b) Satisfaction scores of the two groups.}
 \vspace{-0.2in}
 \label{fig13}
\end{figure}

We collect all subjects' videos and manually extract their operation sequences and the duration of each operation for post-analysis. Figure \ref{fig13}(a) shows the operation sequences of all subjects. GA subjects conduct fewer operations and take less time than GB subjects. ANOVA results show significant effects of knowledge incorporation on completion time ($F$(1, 20) = 247.96; $p$ $\textless$ 0.005) and number of operations ($F$(1,20) = 165.37; $p$ $\textless$ 0.005). In addition, GA subjects always quickly select an attribute to create groups and only adjust CLR 1-2 times.
\\[5pt]
\noindent \textbf{Insight 1.} \emph{The results are in line with our expectations. Knowledge incorporation prompts the formation of clear and class-consistent visual clusters, thus facilitating the group selection of GA subjects. In contrast, the sample overlap hinders the group discovery of GB subjects. 
}
\\[5pt]
\noindent \textbf{Insight 2.} \emph{The only hyperparameter, i.e., CLR, does not affect the usability of our approach. The analyst can start with a small CLR, such as 5\%-10\%, and gradually increase it during the data exploration.}
\\[5pt]
\indent GA Subjects gave higher satisfaction scores than GB subjects, as in Figure \ref{fig13}(b). ANOVA results show the significant difference between the two groups ($F$(1,20) = 110.85; $p$ $\textless$ 0.005) in satisfaction scores. Representative feedback includes:
\\[5pt]
\noindent \textbf{Feedback 1.} \emph{Knowledge incorporation is helpful for pattern discovery. Clear and class-consistent clusters avoid blind sample selection and facilitate the summarization of patterns. We can easily understand which group a selected cluster refers to based on color.}
\\[5pt]
\noindent \textbf{Feedback 2.} \emph{The workflow and visual design are reasonable. We can easily understand the system and learn how to use it. The factor list of PE is practical. The two types of factors are impressive, and I can quickly know the main factors that lead to the cluster formation by looking at the topmost ones.}
\\[5pt]
\indent There are also a few negative comments and questions, as follows: 
\\[5pt]
\noindent \textbf{Feedback 3.} \emph{The tree (KE), although innovative, is not easy to understand. Why not achieve KE with standard controls, such as table and list? ($P_2$) The dataset is small. Whether the approach also suitable for large datasets (more records and dimensions)? ($P_8$, $P_{19}$, $P_{20}$) }
\\[5pt]
We answer these questions. Specifically, the tree structure is for achieving the progressive grouping, and the approach is suitable for large datasets (see experimental results in Section 8.2).

\section{Quantitative experiments}
\label{Quantitative experiments}

We conducted two experiments to demonstrate the performance of the embedding network. In both experiments, we choose SAE (Supervised Autoencoder) \cite{le2018supervised} as a reference.

\subsection{Clustering Accuracy}
We first tested the performance of our network in the downstream clustering task on three labeled open datasets, i.e., MNIST\cite{lecun1998gradient}, CIFAR-10\cite{krizhevsky2009learning}, Libras Movement\cite{nr}. We generated embeddings and clustered the embeddings into groups for each dataset. We counted the percentage of samples clustered into the correct groups as the accuracy. 

Table \ref{tab1} shows the experiment results. An obvious trend is that the accuracy gradually increases as CLRs, which is consistent with our expectation (high CLRs can pull samples of the same class close, see Figure \ref{fig8}). On the other hand, our network has a higher clustering accuracy than SAE on all conditions. Especially on CIFAR-10 and Libras Movement, the accuracy of SAE is low even under higher CLRs. We consider this because we set very few epochs (100). However, our method achieves higher accuracy under the same conditions. The results prove that our network can promote visual structure formation (Section 4) and is more suitable for visual analytics that needs a rapidly converging model to improve the system response speed.

\begin{table}
\renewcommand\arraystretch{1.1}
\small
\centering
\caption{Comparison of our network and SAE in clustering accuracy.}
\label{tab1}
\resizebox{\linewidth}{!}{
\begin{tabular}{ccccccc} 
\toprule
\multirow{2}{*}{\begin{tabular}[c]{@{}c@{}}epochs=100\\DataSet\end{tabular}} & \multicolumn{2}{c}{CLR=10} & \multicolumn{2}{c}{CLR=50} & \multicolumn{2}{c}{CLR=90}  \\ 
\cline{2-7}
                         & SAE  & Ours             & SAE  & Ours             & SAE  & Ours              \\ 
\hline
MNIST                    & 0.22 & 0.54            & 0.45 & 0.69            & 0.88 & 0.97             \\
CIFAR-10                 & 0.25 & 0.33            & 0.28 & 0.51            & 0.29 & 0.80             \\
Libras~Movement          & 0.18 & 0.69            & 0.34 & 0.82            & 0.62 & 0.86             \\
\bottomrule
\end{tabular}
}
\end{table}

\begin{table}
\renewcommand\arraystretch{1.0}
\small
\centering
\caption{Comparison of our network and SAE in training time.}
\label{tab2}
\resizebox{\linewidth}{!}{
\begin{tabular}{ccccccc} 
\toprule
\multirow{2}{*}{\begin{tabular}[c]{@{}c@{}}epochs=100\\$\left| Dimension \right|$ \end{tabular}} & \multicolumn{2}{c}{N=100} & \multicolumn{2}{c}{N=500} & \multicolumn{2}{c}{N=1000}  \\ 
\cline{2-7}
                                                                                & SAE  & Our                & SAE  & Our                & SAE  & Our                  \\ 
\hline
1000                                                                            & 0.53 & 0.43               & 0.91 & 0.77               & 1.29 & 1.13                 \\
5000                                                                            & 1.08 & 0.87               & 2.50 & 2.00               & 3.99 & 3.15                 \\
10000                                                                           & 1.74 & 1.34               & 4.22 & 3.34               & 7.28 & 5.73                 \\
\bottomrule
\end{tabular}
}
\vspace{-0.2in}
\end{table}

\subsection{Training Efficiency}
We finally tested the training speed of the network on synthetic datasets with different numbers of samples and features. Each sample has a multi-dimensional feature (random numbers) and a label. We experimented on an ordinary PC (AMD Ryzen 7 4800H, 16G, GTX1650).

Table \ref{tab2} shows the experiment results. The training time of our network gradually increases with the sample amount (x-axis direction) and embedding width (y-axis direction). Second, our network takes less time than SAE under all conditions, indicating better training efficiency (SAE requires another execution process to compute embeddings after the network training, which is unnecessary for our network).

\section{Conclusion and Future works} 
\label{Conclusion and Future works}
We have presented a novel visual analytics approach to identifying weak patterns prevalent in multi-dimensional data due to dimensionality sparsity. Analysts likely miss these weak patterns during the interactive data exploration since they cannot form clear visual structures in the projection. We have resolved this problem by externalizing tacit human knowledge and incorporating the knowledge into data embeddings by adding a classification loss in the embedding network. The approach combines the strengths of human and machine learning algorithms, achieving data analysis of human-machine hybrid intelligence. Our research may facilitate the emergence of more similar studies incorporating different types of knowledge into data embeddings to facilitate pattern discovery in the projection. 

Scalability is an easily overlooked aspect. Although the embedding network allows for almost real-time responses when handling datasets of the same scales as those used in the case studies, its performance inevitably deteriorates as the rows and columns of the dataset increase. The biggest problem is that increasing columns leads to more neurons in each network layer. It is impossible for a fully-connected layer that integrates thousands of neurons to handle datasets of the same width. A feasible method is to utilize a multi-subnets structure \cite{xie2021exploring} that sets multiple small subnets for handling different parts of dimensions. The structure can reduce the parameter amounts by converting an $M \times M$ layer connection into multiple $N \times N$ ones ($N \ll M$, we assume two adjacent layers have the same number of neurons for simplicity). We plan to replace our single-network structure with a multi-subnet structure and test its performance systematically in the future.

 \acknowledgments{
 This work is supported by the NSFC project (61972278) and Natural Science
Foundation of Tianjin (20JCQNJC01620)}

\bibliographystyle{abbrv-doi}

\bibliography{template}
\end{document}